\documentclass[fleqn,usenatbib,useAMS]{mnras}

\usepackage{graphicx}	
\usepackage{amsmath}	
\usepackage{amssymb}	
\usepackage{multicol}        
\usepackage{bm}		
\usepackage{pdflscape}	

\usepackage{caption}
\usepackage{subcaption}

\usepackage[T1]{fontenc}
\usepackage{ae,aecompl}

\usepackage{newtxtext,newtxmath}
\usepackage{tablefootnote}
\usepackage{array,multirow,graphicx}
 
\usepackage{listings}
\title[JexoSim 2.0]{JexoSim 2.0: End-to-End JWST Simulator for Exoplanet Spectroscopy - Implementation and Case Studies}

\author[S. Sarkar \& N. Madhusudhan]
{
Subhajit Sarkar$^{1}$ 
\thanks{Contact e-mail: \href{mailto:subhajit.sarkar@astro.cf.ac.uk}{subhajit.sarkar@astro.cf.ac.uk}}%
\thanks{School of Physics and Astronomy, Cardiff University, The Parade, Cardiff, CF24 3AA, UK},
Nikku Madhusudhan$^{2}$
\\
$^{1}$School of Physics and Astronomy, Cardiff University, The Parade, Cardiff, CF24 3AA, UK. \\
$^{2}$Institute of Astronomy, University of Cambridge, Madingley Road, Cambridge CB3 0HA, UK
}

\date{Last updated 2020 June 10; in original form 2013 September 5}

\pubyear{2020}

\begin{document}
\label{firstpage}
\pagerange{\pageref{firstpage}--\pageref{lastpage}}
\maketitle

\begin{abstract}
The recently developed JWST Exoplanet Observation Simulator (JexoSim) simulates transit spectroscopic observations of exoplanets by JWST with each of its four instruments using a time-domain approach. Previously we reported the validation of JexoSim against Pandexo and instrument team simulators. In the present study, we report a substantially enhanced version, JexoSim 2.0,  which  improves on the original version through  incorporation of new noise sources, enhanced treatment of stellar and planetary signals and instrumental effects, as well as improved user-operability and optimisations for increased speed and efficiency. 
A near complete set of instrument modes for exoplanet time-series observations is now included. In this paper we report the implementation of JexoSim 2.0 and assess performance metrics for JWST in end-member scenarios using the hot Jupiter HD 209458 b and the mini-Neptune K2-18 b. We show how JexoSim can be used to compare performance across the different JWST instruments, selecting an optimal combination of instrument and subarray modes, producing synthetic transmission spectra for each planet. These studies indicate that the 1.4 \textmu m water feature detected in the atmosphere of K2-18 b using the Hubble WFC3 might be observable in just one transit observation with JWST with either NIRISS or NIRSpec. JexoSim 2.0 can be used to investigate the impact of complex  noise and systematic effects  on the final spectrum, plan observations and test the feasibility of novel science cases for JWST.  It can also be customised for other astrophysical applications beyond exoplanet spectroscopy.  JexoSim 2.0 is now available for use by the scientific community.

\end{abstract}




\section{Introduction}

Since the discovery of the first extra-solar planets a quarter of a century ago \citep{Mayor1995,Wolszczan}, our understanding of planets around other stars has expanded greatly.  We know of the existence of over 4000 exoplanets\footnote{\label{note1}https://exoplanetarchive.ipac.caltech.edu/} covering a diverse range of sizes, masses, temperatures and orbital configurations.  We find that our Solar System is far from typical in the Galaxy.  The discovery of diverse planets in wide-ranging system architectures has challenged planet formation and evolution theories, with a need to understand the factors and conditions that bring about the diversity seen.  To this end, atmospheric spectroscopy characterises exoplanets by obtaining compositional, structural and dynamic information of their atmospheres \citep{Madhusudhan2019}. The atmospheric compositions can in turn provide important insights into planetary formation and evolution mechanisms. Spectroscopy is also key in the search for biosignature gases indicative of life \citep{Seager2014, Schwieterman2018}.  Atmospheric spectroscopy of exoplanets has been achieved through the transit-based techniques of transmission and eclipse spectroscopy   \citep{Seager2000,Charb2002,Grillmair2008, Deming2013} , as well as direct imaging \citep{Konopacky1398, Janson_2010}. The latter, while not constrained by the need for transits, is currently limited to studying young gas-giant planets at wide separations from their host stars.  

Transmission spectroscopy depends on  wavelength-dependent absorption of stellar light by atmospheric atomic and molecular species, as the planet transits in front of the star \citep[e.g.,][]{Charb2002, Sing2016, Diamond-Lowe2020}.  This technique probes the high altitude atmosphere of the planet day-night terminator, returning a spectrum consisting of $(R_p/R_s)^2$ with wavelength, where $R_p$ is the apparent planet radius and $R_s$ is the star radius.  It has been applied in the visible and near-infrared wavelength ranges where the star is brighter, allowing higher signal-to-noise, thereby making detection of the spectral features of key chemical species more accessible at these wavelengths.  Eclipse spectroscopy depends on the occultation of the planet's reflection and emission flux as it passes behind the star (secondary eclipse).  The eclipse depth variations with wavelength give the spectrum of $F_p/(F_s+F_p)$, where  $F_p$ is the planet flux and $F_s$ is the stellar flux.  Eclipse spectroscopy \ provides constraints on the temperature structure and composition of the dayside atmosphere  \citep[e.g.,][]{Stevenson2014, Line2013, Madhusudhan2018}. 

The Hubble space telescope has been used to great effect in obtaining transmission spectra of exoplanets using instruments such as STIS, NICMOS, ACS and WFC3 \citep{Charb2002, Swain2008, PontACS2008, SingSTIS2011, Kreidberg2014} however the passbands of the spectra obtained are narrow (e.g. 1.1-1.7 \textmu m for the Hubble WFC3) so that only small fraction of the spectrum is revealed.  The Spitzer space telescope has also been used to obtain transmission and emission spectra, but these consisted of just a few photometric bands  \citep{AgolSpitzer_2010, Knutson_2011, Deming2020}  . Ground-based facilities are hampered by
the infra-red absorption windows of the Earth's atmosphere, as well as having to contend with atmospheric scintillation and turbulence effects.  
Wider wavelength coverage is desirable to reduce degeneracies and overcome impact of clouds \citep{Barstow2015, Kreidberg2014, Sing2016, Madhusudhan2019, Welbanks2019}.
The next generation of transmission and emission spectra will be obtained from space-based observatories with wide wavelength coverage such as the James Webb Space Telescope (JWST) \citep{Gardner2006} and Ariel \citep{Tinetti2018}, greatly improving the quality and quantity of exoplanet spectra obtained.   

JWST is planned for launch in 2021.  All four instruments on board are able to perform transit spectroscopy of exoplanets. These are the Near Infrared Imager and Slitless Spectrograph (NIRISS) \citep{Doyon2012}, the Near Infrared Camera (NIRCam) \citep{Beichman2012}, the Near Infrared Spectrograph (NIRSpec) \citep{Ferruit2014}, and the Mid-Infrared Instrument (MIRI)  \citep{Rieke2015}.  Combined, these instruments provide a wavelength coverage from 0.5 to  28.3 \textmu m . 
JWST's 6.5m primary mirror will be largest to date in space, reducing photon noise to unprecedented levels. 
Reduced photon noise however could mean that other noise sources and systematics become more important and impact the data.  For JWST therefore, it is particularly important to consider additional sources of noise or systematic trends that could possibly confound the detection of small signals such as exoplanet atmospheric signatures afforded by its low photon noise.  Proper accounting of all errors is also crucial when establishing the final experimental uncertainties of a study including the impact of correlated noise.  The size of the final error bars on the planet spectrum impacts the uncertainties in spectral retrievals \citep[e.g.][]{Madhusudhan2009, Line2013, Waldmann2015, Madhusudhan2018} and the scientific conclusions of a study.  Simulators 
that model the instrumental and astrophysical noise, as well as systematics, during an observation can help to assess the performance of JWST.  They can be used in the planning and optimisation of observational and data reduction strategies, and for testing the feasibility of specific science cases. 

Since transit spectroscopy is a time-domain technique it is particularly vulnerable to time-dependent systematic trends from both the instrument, such as detector persistence \citep{Berta2012}, or from astrophysical sources, such as spot occultations and stellar variability \citep{Rackham2018, Pont2006}.  In addition, time-correlated noise, e.g from pointing jitter or stellar granulation or pulsation \citep{Sarkar2018}, can inflate the final uncertainties and needs to be properly accounted for \citep{Pont2006}.  Most simulators for JWST do not model the time-domain directly, but estimate the final noise on the transit depth through a static calculation.
This approach leads to rapid computations which is ideal for assessing large sets of planets, however they lack the functionality to model time-correlated noise or time-dependent systematic trends.
Pandexo \citep{Batalha2017} simulates all four JWST instruments giving signal-to-noise (SNR) estimates based on set of complex computations using the Pandeia engine \citep{Pontoppidan}.  While Pandexo generates 2-D detector images allowing it to model spatially-correlated noise sources, it does not produce a frame-by-frame time-domain simulation which limits its ability to model complex time-dependent processes.
Simulators that have been produced by individual JWST instrument teams include the NIRISS SOSS 1-D\footnote{http://maestria.astro.umontreal.ca/niriss/simu1D/simu1D.php} \citep{Beichman2014} and 2-D simulators\footnote{https://github.com/jasonfrowe/JWSTNIRISS},
 a 1-D simulator developed for NIRCam and MIRI \citep{Greene_2016},  PyNRC for NIRCam \footnote{https://github.com/JarronL/pynrc}  \citep{Leisenring2018}, the NIRSpec instrument performance simulator \citep{Piqueras2008} and the NIRSpec Exoplanet Exposure Time Calculator (NEETC) \citep{Nielsen2016}. To our knowledge these do not produce frame-by-frame time-domain simulations.

We developed the first version of JexoSim (the JWST Exoplanet Observation Simulator) \citep{Sarkar2020} to fill this gap in the range of simulators available for JWST. Uniquely, JexoSim can simulate exoplanet transit observations using all four JWST instruments using a frame-by-frame time-domain approach, generating 2-D image time-series akin to an actual observation. It can model and capture the effects of time-correlated noise and systematic trends. JexoSim can also capture 2-D spatial- and wavelength-dependent systematics, and the potentially complex interaction between spatial and temporal processes, allowing measurement of their ensemble impact directly on the planet spectrum.  

In the previous work \citep{Sarkar2020}, we described the JexoSim algorithm in detail and validated the simulator against Pandexo and instrument team simulators.  In the current work, we report an enhanced version, JexoSim 2.0,  with new noise sources, added instrumental effects including quantum yield and inter-pixel capacitance and enhanced fidelity in the stellar and planetary signal processing algorithm.  We have made it  faster, more versatile and easier to use.  User-operability has been greatly improved with several architectural changes including a simplified user interface and installation procedure, together with more up-to-date and reduced number of dependent packages.  In additional, a number of optimisations have been implemented to improve speed, reduce memory pressure, and enable the simulation of observations with very large numbers of integrations.  We now include a near complete set of JWST instrument modes designated for exoplanet time-series observations.  
With these enhancements JexoSim 2.0 is now available to the community\footnote{https://github.com/subisarkar/JexoSim }. 

In this paper, we describe the upgrades made in JexoSim 2.0 and provide a guide to its implementation.  We then use the simulator to assess JWST performance metrics for specific case studies.  We focus on exoplanet transmission spectroscopy using two end-member scenarios: the hot Jupiter HD 209458 b and the mini-Neptune K2-18b.  We perform simulated observations using all four JWST instruments and demonstrate different simulation recipes. These include using Allan deviation analysis and Monte Carlo simulations to obtain precision on  transit and eclipse depths  and a noise budget analysis.  We use JexoSim to assess which  instrumental configurations are optimal for these end-member scenarios.

\section{JexoSim 2.0}

\begin{figure*}
\begin{center}
    \includegraphics[trim={0 0 0 0}, clip,width=1.0\textwidth]{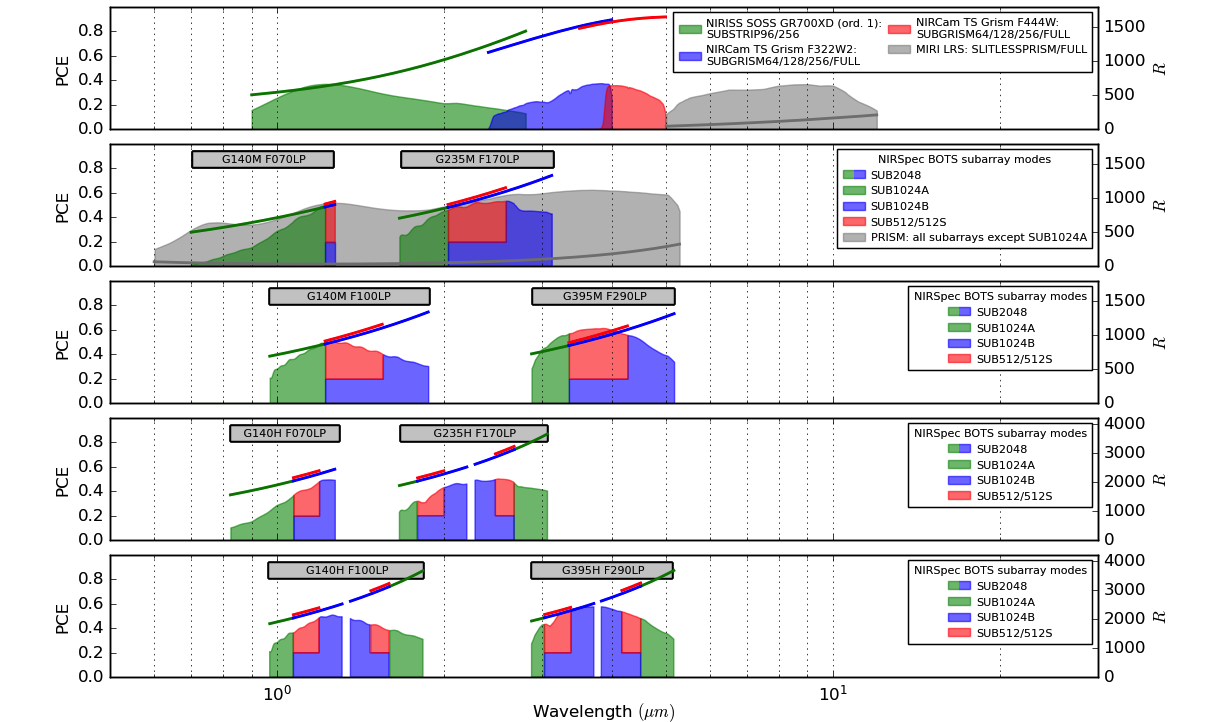}
\caption[]{JexoSim 2.0 available instrument configurations shown with their photon-conversion efficiency (PCE) (upper bounds of shaded areas) and spectral resolving power ($R$) (floating lines) versus wavelength.  Nominal wavelength ranges for each mode are shown. In NIRSpec the SUB512 and SUB512S modes have wavelength ranges which are subsets of the SUB1024B mode, hence where they overlap in range the colours are split between the modes (SUB512/SUB512S in red, SUB1024B in blue).  SUB2048 encompasses the full wavelength range of each NIRSpec mode, i.e. the combination of green and blue areas.  
JexoSim 2.0 uses instrument-specific transmissions and quantum efficiency files from the Pandeia database \citep{Pontoppidan}, producing the final PCEs shown. The $R$ power information is also derived from Pandeia. Where modes overlap, the $R$ powers are slightly offset for clarity. 
SOSS = Single Object Slitless Spectroscopy. BOTS = Bright Object Time Series. TS Grism = Time Series Grism.  LRS = Low Resolution Spectrometer. }
\label{Fig:PCE}
\end{center}
\end{figure*}

JexoSim 2.0 is a time-domain simulator of exoplanet transit spectroscopy using JWST and it four instruments, designed to test and validate specific and novel science cases for JWST observation.  Both primary and secondary transits can be simulated, as well as out-of-transit (OOT) observations. Time-series images are generated representing the sequence of non-destructive reads (NDRs) per integration.  The images can be processed using an included pipeline for immediate results, or stored in FITS format for later analysis.  Simulated planet spectra with error bars can be generated, which can be used to test the detectability of atmospheric features, e.g. through spectral retrieval studies.  Alternately signal and noise information from the OOT stellar signal can be extracted.   Nearly all JWST instrument configurations dedicated to time-series exoplanet spectroscopy are now included in JexoSim 2.0, with their available subarray modes (Figure \ref{Fig:PCE}),  although the MIRI Medium Resolution Spectrograph (MRS) mode is currently not simulated. In NIRISS only the first order spectrum from the GR700XD grism is currently simulated.   

Next, we describe the upgrades made in JexoSim 2.0 in architecture,  algorithm, dependencies and optimisations.

\subsection{Architecture}
\begin{figure*}
\begin{center}
    \includegraphics[trim={0 0 0 0}, clip,width=0.9\textwidth]{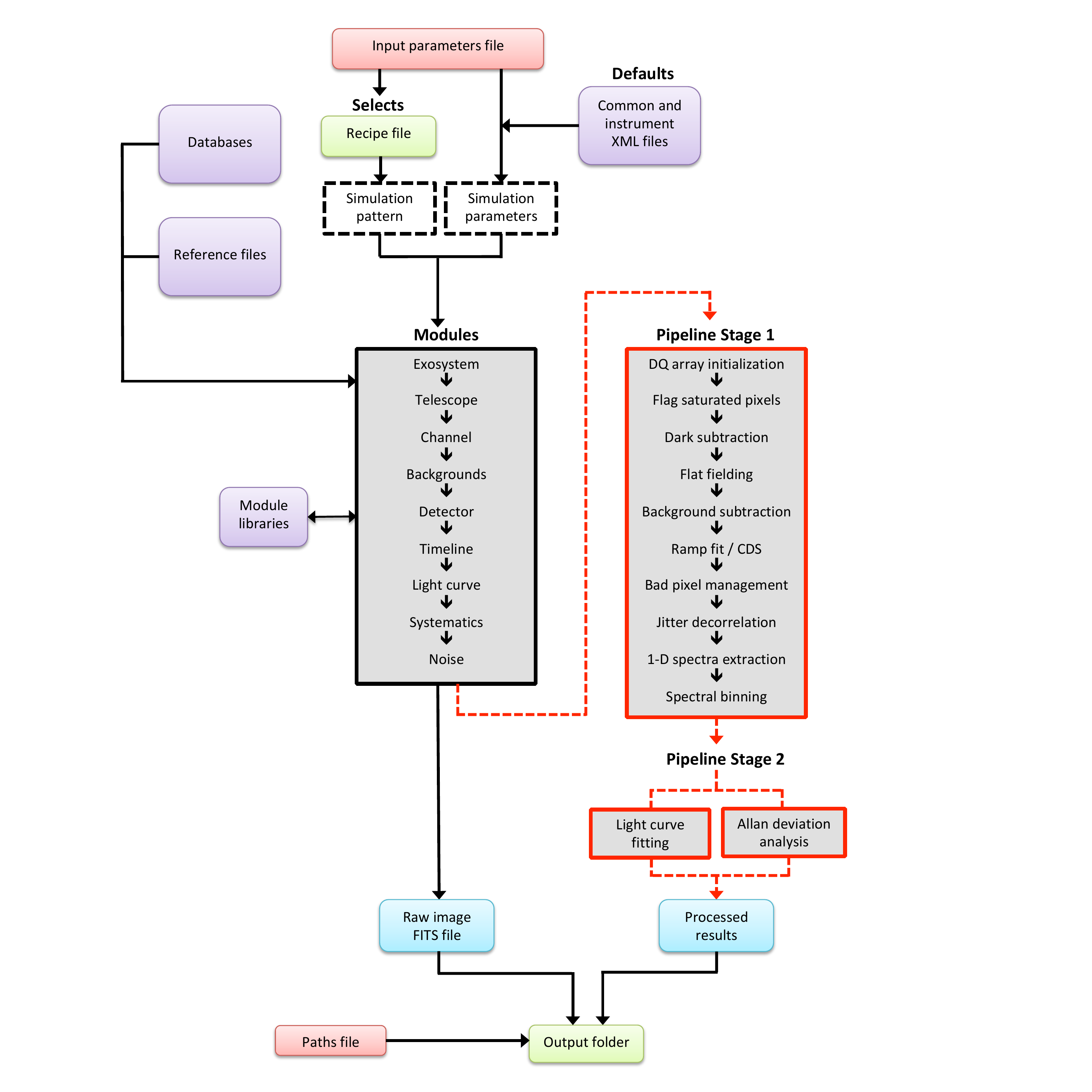}
\caption[]{Overview of JexoSim 2.0 architecture.  A new 
`input parameters file' provides the user-interface and gives control over the simulation.  Recipe files direct the pattern of the simulation, while XML files hold instrumental parameters, links to reference files and a default set of simulation parameters.  The simulation proceeds through a sequence of `modules'.  These are described in \cite{Sarkar2020}, together with the processing algorithm. Modifications to the algorithm in JexoSim 2.0 are illustrated in Figure \ref{Fig: algo}. The modules now link to dedicated libraries.  The final image stack is stored as FITS file or processed via the JexoSim pipeline. }
\label{Fig:architecture}
\end{center}
\end{figure*}

\begin{figure*}
\begin{center}
    \includegraphics[trim={1cm 2.5cm 4cm 1.5cm}, clip,width=1.0\textwidth]{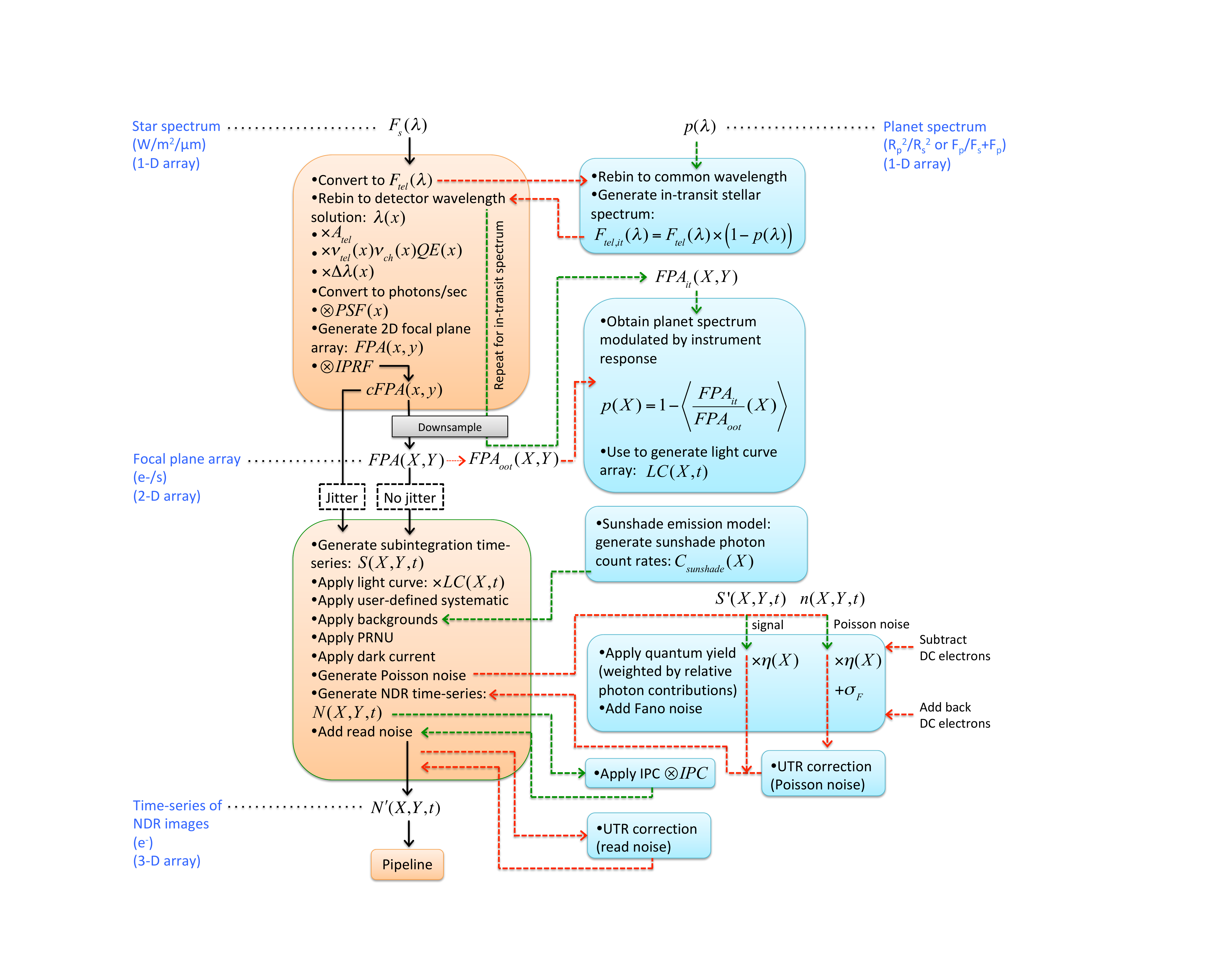}
\caption[]{
Simplified representation of the updated JexoSim algorithm. Right-hand side boxes (blue) show main changes in the algorithm introduced in JexoSim 2.0. Left-hand side boxes (orange) mostly follow the original algorithm, the details of which are completely described in \cite{Sarkar2020}. $X$ and $Y$ are the spectral and spatial coordinates of whole pixels on the focal plane detector array, $FPA(X,Y)$.  An oversampled grid, $FPA(x,y)$, is initially generated (for Nyquist-sampling of the PSF and jitter simulation), which has subpixel coordinates $x$ and $y$. 
$A_{tel}$= telescope aperture area. $\nu_{tel}$= telescope transmission.$\nu_{ch}$= instrument channel transmission. $\Delta \lambda$= wavelength span across subpixel in spectral direction.  $IPRF$= 2-D intra-pixel response function. $cFPA(x,y)$= oversampled focal plane array convolved with intra-pixel response function.  Other abbreviations are described in the figure or text.  }
\label{Fig: algo}
\end{center}
\end{figure*}

The architecture of JexoSim 2.0 is outlined in Figure \ref{Fig:architecture}. A core sequence of modules is supported with databases and reference files, with various inputs and outputs to the system including an optional pipeline.  Here we briefly discuss these components.

\subsubsection{Modules}
JexoSim 2.0 retains from the original version the sequence of modules that executes the processing algorithm \citep{Sarkar2020}.  The modules simulate the flow and modulation of the signal from the star to the detector followed by the the generation of a time-series image stack of non-destructive reads (NDRs).  The modules however now refer to dedicated libraries that hold associated functions, while the modules themselves control the calling of these functions. This organisation will aid future augmentation allowing for easier upgrades and addition of new functions.

\subsubsection{Inputs}
To improve user operability, full control over the simulation is now achieved through a new text-based `input parameters file' (Figure \ref{Fig:architecture}).  This file gives the user wide control over the simulation parameters and can be edited or duplicated to run particular simulation scenarios.  A new `paths file' (also a text file) allows the user to specify the path of the output folder for results.  As in the previous version, default simulation and instrument parameters are held in XML-based configuration files which link to specific reference files (e.g. transmissions, wavelength solutions etc.).  A new \texttt{params} class mediates the interaction of the input parameters file with the rest of the code modifying default values held in the XML files. A new set of XML files has been added for extra-solar system (`exosystem') parameters. When an exoplanet is first referenced from the planet database (see Section \ref{sec:dependencies}), an XML file containing the system parameters is generated and then re-used in subsequent simulations of the same system.  A new set of files called `recipe' files have been added which control patterns of simulation, types of results and outputs.  The code will select a particular recipe depending on choices made in the input parameters file.

\subsubsection{Outputs}
The time-series of NDR images is the final output of the signal processing algorithm.  This can be packaged into FITS format for processing by an external pipeline.  Alternately, the user can choose to process the images through the included JexoSim pipeline for immediate results.  The main steps in this pipeline are shown in Figure \ref{Fig:architecture} and described in detail in \cite{Sarkar2020}.  If processed through the pipeline, the output files are produced in .pickle format (containing processed data) and .txt format (containing parameters from the input parameters file or derived from the simulation itself).
All output files are dumped to the specified output folder.

\subsection{Algorithm}
The original signal processing algorithm (the sequence of calculations executed in the modules) was described in detail in \cite{Sarkar2020} and illustrated in Fig. 1 of that paper. In the new version, the algorithm code has been refactored for efficiency and the entire JexoSim code updated from Python 2 from Python. Modifications to the algorithm are described below.  A summary of the algorithm and changes made in JexoSim 2.0 is illustrated in Figure \ref{Fig: algo}.

\subsubsection{Stellar and planetary spectra}
\label{stellar signal}
To account for resolution-linked bias (RLB) \citep{Deming2017}, JexoSim 2.0 has a modified treatment of the star and planet spectra. The bias refers to a mismatch between the real transit depth in a spectral bin and that predicted using a transit spectrum model if the model utilises insufficient spectral resolution.
If the planet spectrum, $p(\lambda)$, is binned (or convolved with the PSF) \textit{prior} to it modulating the out-of-transit stellar spectrum, $F_s(\lambda)$, this bias is likely to result, and the fidelity of the simulation reduced.  The bias increases when absorption lines from the star and planet overlap.  To maximise fidelity by minimising the RLB effect, JexoSim now multiplies the planet spectrum into the stellar spectrum early in the algorithm while both are at their highest resolution (Figure \ref{Fig: algo}). The `in-transit' stellar spectrum, is next obtained: $F_{s, it}(\lambda) = F_s(\lambda)[1-p(\lambda)]$.
The in-transit and out-of-transit stellar spectra are then treated in exactly the same way through a sequence of steps ending with the production of a 2-D focal plane array for the out-of-transit signal, $FPA_{oot}(X,Y)$, and the in-transit signal, $FPA_{it}(X,Y)$, where $X$ and $Y$ are pixel coordinates. 
Figure \ref{Fig: algo} summarises these steps and a full description is given in \cite{Sarkar2020}.  We obtain the ratio of these 2-D arrays, $FPA_{it}/FPA_{oot}$, and then  find the average ratio per pixel column, $\left< FPA_{it}(X)/FPA_{oot}(X) \right>$.  This returns a 1-D array.  Subtracting this 1-D array from 1, gives $p(X)$, the planet spectrum correctly modulated by the instrument response (Figure \ref{Fig: algo}).  $p(X)$ is used as the input to generating the light curve 2-D array, $LC(X,Y,t)$, where $t$ corresponds to the time of a subintegration\footnote{We use the term `subintegration' to define as the count accumulated between two successive NDRs (i.e. the difference between them). Co-adding the subintegrations of an integration subsequently generates the ramped NDRs for that integration.}.  $FPA_{oot}$ is used to generate the subintegration time-series, $S(X,Y,t)$, which is modulated by the light curves, systematics, backgrounds and noise (Figure \ref{Fig: algo}).

To minimise RLB the user can supply both the planet spectrum and the star spectrum at very high resolution. JexoSim allows for the user to supply their own stellar and/or planet spectra through the options in the input parameters file (Table \ref{tab: input parameters}) (select \texttt{file} for options 
\texttt{planet\textunderscore spectrum\textunderscore model} and \texttt{star\textunderscore spectrum\textunderscore model}).  As mentioned in Section \ref{sec:dependencies}, JexoSim uses a default stellar Phoenix BT-Settl spectrum database.  For star parameters, 3000 K, log$g$ 5.0, Fe/H = 0, this gives a spectrum with average $R$-power of $\sim$ 390000 between 0.05-5.5 \textmu m.  For the same star and wavelength range, high-resolution Phoenix spectrum by \cite{Husser2013} \footnote{https://phoenix.astro.physik.uni-goettingen.de/?page\textunderscore id=15} give an average $R$-power of $\sim$ 610000, and can be used in JexoSim 2.0 by selecting for a user-supplied stellar file as described above \footnote{These need to be first repackaged as text files with the first column giving the wavelength in \textmu m, and the second column the stellar flux density in W/m$^2$/\textmu m. The wavelength grid of these files extends from 0.5-5.5 \textmu m. JexoSim will return zero value for the stellar flux density outside this range, and thus these spectra are not suitable for MIRI LRS.}.  
After selecting the star and planet spectra, the code must rebin these to a common wavelength grid (that of the spectrum with the lower resolution) to allow for array multiplication (Figure \ref{Fig: algo}).
Thus if the initial planet spectrum is provided at much lower $R$ than the stellar spectrum, there will be some inevitable loss of fidelity (re: the RLB effect) due to rebinning at this stage. This can be minimised by providing both planet and star spectra at high resolution.  

As an alternative to the user providing a specific planet spectrum as input, in JexoSim 2.0 we have added the capability to use the fully scalable ATMO model grid of transmission spectra produced by \cite{Goyal2018}\footnote{https://drive.google.com/drive/folders/1ZFbkPdqg37\_ Om7ECSspSpEp5QrUMfA9J} for gaseous planets, as an option for the input planet spectrum.  Alternately the user can select a `simple' flat planet transmission spectrum based on $(R_p/R_s)^2$ or a blackbody-based emission spectrum.  
Table \ref{tab: input parameters} shows how to choose between these options in the input parameters file.

\subsubsection{Backgrounds}
JexoSim 2.0 uses the same zodiacal light and optical surfaces emission models developed for the previous version \citep{Sarkar2020}.  In addition to these,  JexoSim 2.0 now includes the thermal emission from the sunshield as a background source (with an associated Poisson noise contribution).  We utilise the publicly-available plot of equivalent in-field sky irradiance vs wavelength for the sunshield from the JWST User Documentation Background Model webpage\footnote{Fig. 2 from https://jwst-docs.stsci.edu/jwst-observatory-functionality/jwst-background-model}.  This is sampled using WebPlotDigitizer \citep{Rohatgi2020} and converted from MJy/sr to W/m$^2$/\textmu m/sr.  A 7th-order polynomial is fitted to the sampled points. This polynomial is used to obtain the spectral flux density at the wavelength solution of the instrument above 7 \textmu m.  Below 7 \textmu m, the flux is assumed to be negligible and set to zero.  It is then processed in the same way as the zodi background, described in \citep{Sarkar2020}.  This results in a 1-D array of photon count-rate per $X$ pixel, $C_{sunshade}(X)$, which is applied to the image time-series  as a background and contributes to the total Poisson noise (Figure \ref{Fig: algo}).  The list of current backgrounds is given in Table \ref{table: summary of noise table}.

\subsubsection{Noise and quantum yield}
Noise sources in the original version of JexoSim \citep{Sarkar2020} included: star (source photon noise), dark current, read-out noise, zodiacal reflected light and thermal emission, optical surfaces emission and pointing jitter.  In JexoSim 2.0 we have added Poisson noise from the sunshade background and Fano noise (Figure \ref{Fig: algo}).  Table \ref{table: summary of noise table} lists the current noise sources simulated.

Fano noise is a type of photon-transfer noise \citep{Janesick2007}, that results from variation in the number of electron-hole pairs in the detector semi-conductor material per absorbed photon. The standard deviation is given by:
\begin{equation}
    \sigma_f (\lambda) = \sqrt{F (\lambda) \eta(\lambda)}
\end{equation}
where $F$ is the Fano factor, and $\eta$ is the quantum yield.  The Fano factor is the ratio of the variance in the number of electrons generated to the average number of electrons generated per photon.  We use the two-level approximation (valid for $\eta$ $\sim$ 1 to 2) for the Fano factor from \cite{McCullough2008}, where $F = [(3 \eta - 2) - \eta^2] / \eta$.  Since both $\eta$ and $F$ are wavelength-dependent, the Fano noise is also.  For $\eta$ = 1, we obtain a Fano noise of zero e$^-$.  In the algorithm (Figure \ref{Fig: algo}), both stellar and background photons are added and Poisson variations to the photon numbers generated before the Fano noise is added.
The wavelength-dependent quantum yield, $\eta(\lambda)$, is obtained for each instrument channel from the Pandeia database.  For the point source we simply rebin this to the wavelength solution for each $X$ pixel column, to give $\eta_*(X)$ for the stellar photons.  For diffuse backgrounds, since a broad range of wavelengths will fall on each pixel (depending on the slit size used), a weighted quantum yield for each $X$ pixel, $\eta_w(X)$ is calculated separately for each diffuse source. 
The quantum yield for each pixel on each subintegration, $\eta(X,Y,t)$, is found from the weighted average of $\eta_*(X)$ and $\eta_w(X)$ for each diffuse source, where the weights are the relative photon contributions on the pixel.  For computational efficiency this weighted quantum yield is multiplied into all the signal and noise photons at a single step in the algorithm (Figure \ref{Fig: algo}).
The Fano noise is then simulated. For each pixel on each subintegration, $\sigma_F(X,Y,t)$ = $\sqrt{SF\eta(X,Y,t)}$, where $S$ is the number of photons on a pixel. This is used to apply a random Gaussian variation in the electron count.  For computation efficiency, the dark current is added prior to the generation of Poisson noise (so that Poisson noise generation is performed once), and thus its contribution to the signal and Poisson noise is subtracted prior to applying quantum yield and Fano noise (since dark current does not result from interacting photons), with the contributions added back after these steps.

\begin{table}
\
\begin{center}
\caption{Summary of noise sources, backgrounds, systematics and instrumental effects included in JexoSim 2.0.}
\label{table: summary of noise table}
\begin{tabular}{lll} 
\hline
\hline
\multicolumn{1}{l}{Effect} &
\multicolumn{1}{l}{Type} &
\multicolumn{1}{l}{New in} 
\\
\multicolumn{1}{l}{} &
\multicolumn{1}{l}{} &
\multicolumn{1}{l}{JexoSim 2.0} 
\\
\hline
\multicolumn{1}{l}{Noise} &
\multicolumn{1}{l}{Star (source photon noise)} &
\multicolumn{1}{l}{}
\\
\multicolumn{1}{l}{sources} &
\multicolumn{1}{l}{Dark current} &
\multicolumn{1}{l}{}
\\
\multicolumn{1}{l}{} &
\multicolumn{1}{l}{Read-out noise} &
\multicolumn{1}{l}{}
\\
\multicolumn{1}{l}{} &
\multicolumn{1}{l}{Zodiacal light} &
\multicolumn{1}{l}{}
\\
\multicolumn{1}{l}{} &
\multicolumn{1}{l}{Optical surfaces emission} &
\multicolumn{1}{l}{}
\\
\multicolumn{1}{l}{} &
\multicolumn{1}{l}{Sunshade emission} &
\multicolumn{1}{c}{$\checkmark$}
\\
\multicolumn{1}{l}{} &
\multicolumn{1}{l}{Fano noise} &
\multicolumn{1}{c}{$\checkmark$}
\\
\multicolumn{1}{l}{} &
\multicolumn{1}{l}{Pointing jitter} &
\multicolumn{1}{l}{ }
\\
\hline
\multicolumn{1}{l}{Background} &
\multicolumn{1}{l}{Dark current} &
\multicolumn{1}{l}{}
\\
\multicolumn{1}{l}{sources} &
\multicolumn{1}{l}{Zodiacal light} &
\multicolumn{1}{l}{}
\\
\multicolumn{1}{l}{} &
\multicolumn{1}{l}{Optical surfaces emission} &
\multicolumn{1}{l}{}
\\
\multicolumn{1}{l}{} &
\multicolumn{1}{l}{Sunshade emission} &
\multicolumn{1}{c}{$\checkmark$}
\\
\hline
\multicolumn{1}{l}{Systematics /} &
\multicolumn{1}{l}{Optical transmissions} &
\multicolumn{1}{l}{}
\\
\multicolumn{1}{l}{instrumental} &
\multicolumn{1}{l}{Wavelength solutions and spectral traces} &
\multicolumn{1}{l}{}
\\
\multicolumn{1}{l}{effects} &
\multicolumn{1}{l}{PSF} &
\multicolumn{1}{l}{}
\\
\multicolumn{1}{l}{} &
\multicolumn{1}{l}{PRNU} &
\multicolumn{1}{l}{}
\\
\multicolumn{1}{l}{} &
\multicolumn{1}{l}{Intra-pixel sensitivity} &
\multicolumn{1}{l}{}
\\
\multicolumn{1}{l}{} &
\multicolumn{1}{l}{Inter-pixel capacitance} &
\multicolumn{1}{c}{$\checkmark$}
\\
\multicolumn{1}{l}{} &
\multicolumn{1}{l}{Wavelength-dependent QE} &
\multicolumn{1}{l}{}
\\
\multicolumn{1}{l}{} &
\multicolumn{1}{l}{Quantum yield} &
\multicolumn{1}{c}{$\checkmark$}
\\
\multicolumn{1}{l}{} &
\multicolumn{1}{l}{Pointing jitter} &
\multicolumn{1}{l}{}
\\
\multicolumn{1}{l}{} &
\multicolumn{1}{l}{User-defined systematic} &
\multicolumn{1}{c}{$\checkmark$}
\\
\hline
\hline
\end{tabular}
\end{center}
\end{table}

\subsubsection{Systematics and instrumental effects}
The list of systematics and instrumental effects currently included are given in Table \ref{table: summary of noise table}.
As in the previous version, photo-response non-uniformity (PRNU) is included as a grid of pixel gain variations.  Intra-pixel sensitivity variations can also be modelled using an intra-pixel response function described in \cite{Sarkar2020}.  In the new version, we have added inter-pixel capacitance (IPC) \citep{Giardino2012} for NIRISS, NIRCam and NIRSpec detectors.  This is applied as a 2-D convolution in the algorithm (Figure \ref{Fig: algo}). 
JexoSim 2.0 allows for the application of an externally-generated user-defined systematic grid, as a 2-D array which modulates the signal in time and wavelength dimensions. To use this function, inside the  \texttt{JexoSim/data/systematic\textunderscore models/} folder must be placed a folder named for the specific systematic which contains text files for the systematic grid (containing the proportional change in signal), wavelength and time.  The input parameters file option (Table \ref{tab: input parameters}) \texttt{sim\textunderscore use\textunderscore systematic\textunderscore model} must be set to 1, and the name of the systematic model folder entered under \texttt{sim\textunderscore systematic\textunderscore model\textunderscore name}.

There may be additional noise sources and systematics not yet represented in JexoSim. Notably we have not yet added the spatially-correlated read noise effect for NIR detectors \citep{Rauscher2017}.  However, JexoSim has a versatile framework where noise sources and systematics can be added in the future, especially as our understanding of the instruments improve during operations.  Astrophysical effects such as the impact of 
star spot effects, stellar pulsation and granulation, transits of exomoons or ring systems, and phase-dependent modulations
can also be investigated using JexoSim by interfacing with external models directly or through the generation of time- and wavelength-dependent flux variation grids from such models which can then be used to modulate JexoSim light curves.

\subsection{Dependencies}
\label{sec:dependencies}
To reduce issues during installation and allow for better future upkeep, we have changed and reduced the dependencies in JexoSim 2.0.The main dependent packages are: PyTransit \citep{Hannu2015}, Astropy \citep{astropy:2013,astropy:2018}, Scipy \citep{2020SciPy-NMeth}, Numpy \citep{2020NumPy-Array}, Pandas \citep{pandas-mckinney-proc-scipy-2010}, Numba \citep{Numba} and Matplotlib \citep{matplotlib}. A dedicated setup process has been developed which imports each of these together with any sub-dependencies.  A number of publicly-available databases are used with JexoSim 2.0.  As in the previous version, the Pandeia database \citep{Pontoppidan}  is used to obtain transmissions and wavelength solutions for each instrument, and a Phoenix BT-Settl database is used for model star spectra \citep{Allard2012}\footnote{\label{note2}https://Phoenix.ens-lyon.fr/Grids/BT-Settl/CIFIST2011\_2015/FITS/}.  We also use values from Pandeia for the instrument dark current, read noise, pixel full-well capacity values, IPC kernels, and quantum yield.  Other instrument parameter values are obtained frrom \cite{Ressler2015}, \cite{NIRISS2019}, \cite{NIRCam2019}, \cite{NIRSpec2019} and \cite{MIRI2019}.  As previously, WebbPSF \citep{Perrin} is used to generate the point-spread-functions (PSFs) for each instrument, and  limb-darkening coefficients (LDCs) are generated using ExoTETHyS \citep{Morello_2020}. Both the PSFs and LDCs are pre-calculated.    
New in JexoSim 2.0 is the use of the NASA Exoplanet Archive$^{\ref{note1}}$ `Planetary Systems' table as the database for extra-solar system parameters. Since this table provides results from multiple studies for a given planet, JexoSim first chooses values from the study which has the most complete set of parameter data. Remaining missing parameters are then selected from studies in order of most recent publication date. In rare cases where a parameter is completely missing, the user will be prompted to enter this manually (this only needs to be done once since the resulting exosystem XML file will be stored for later use). 

\subsection{Optimisation}

The simulation of pointing jitter described in \cite{Sarkar2020} represented one of the slowest sections of the JexoSim algorithm, requiring repetitive sequential shifting of the image array at a subpixel level.  In the previous version we used the \texttt{scipy.weave} function to interface with a section of code written in C that performed these shifts and improved speed.  However, with the change to Python 3, this function can no longer be used.  We use instead the Python optimisation package \texttt{Numba}, combined with its own parallelisation function for this part of the algorithm.  This combination reduces jitter code processing times by an order of magnitude compared to standard Python.  In addition we implement splitting the full image array into longitudinal sections, jittering each section (with the same jitter sequence) and then recombining the sections. This leads to an additional speed improvement by reducing the array size used in the jitter function. 

In some instrument modes it is possible to `crop' the image array in the `Noise' module (Figure \ref{Fig:architecture}) to a smaller size, reducing the overall size of the final time-series array  while preserving the full spectrum. The backgrounds are sampled at the edges of the image prior to cropping and later used for background subtraction in the data pipeline.  Reducing the array size improves speed of the simulation and pipeline processing.
Some simulations may produce very large numbers of NDRs (e.g. many thousands) reducing speed of the simulation and potentially overwhelming computer memory. This is particularly an issue for bright targets that have short integration times, targets with long overall observing times, and instrument modes with short frame times.  

JexoSim 2.0 currently simulates only `fast' read out modes where the number of frames per group ($m$) equals 1: MIRI (FAST),  NIRSpec (NRSRAPID), NIRISS (NISRAPID) and NIRCam (RAPID).  For NIRCam, 1-output and 4-output readout patterns are offered, each of which has a different frame time.
JexoSim can operate using either correlated-doubling-sampling (CDS) or up-the-ramp (UTR) modes. In the latter, a slope is fitted in the pipeline to the NDR ramp in each integration cycle. This reduces read noise compared to CDS, but slightly increases Poisson noise \citep{Rauscher2007}. The user can choose to simulate the ramps by generating every NDR directly.  However, to increase speed and reduce memory pressure, an alternative is offered where only the first and last NDRs are actually generated per integration cycle but the noise is adjusted so that when CDS (last-minus-first subtraction) is performed, the resulting Poisson noise and read noise are at the level expected for a ramp with $n$ NDRs, where  $n$ is the number of groups per integration. 
This is achieved by adjusting the total Poisson noise (and Fano noise) on each NDR by a factor of $\sqrt{ 6(n^2+1)/5n(n+1)} $, and the read noise by a factor of $ \sqrt{6(n-1)/n(n+1)}$ (Figure \ref{Fig: algo}).  We use this alternative mode in all the simulation examples given in this paper.  

For full transit simulations where the number of NDRs is very large (many thousands), potential exists to overwhelm computer memory and slow simulation and pipeline running time due to the very large array size produced.  To mitigate this, when the number of NDRs becomes very large, the timeline of images is split into consecutive segments.  Each segment is processed through stage 1 of the pipeline (Figure \ref{Fig:architecture}) producing spectrally-binned light curve segments which are later  joined together to produce the final light curves.  These are then processed in stage 2 of the pipeline where light curve model fitting takes place to extract the planet spectrum.  

To exclude background pixels which may contribute to noise, an extraction aperture (aperture mask) is applied over the integration image in the pipeline. However the aperture may also reduce signal and increase spatial jitter noise if too narrow. We have added a new function that auto-selects an optimal aperture width which maximises the  stellar signal-to-noise ratio (SNR). If this option is selected, the code will rapidly trial different aperture widths on a sample segment of the image time series, and then selects the width that returns the highest average SNR over all wavelengths.

\section{Implementation}

\subsection{Installation}

JexoSim 2.0 is hosted on Github and is installed as follows.  First a dedicated virtual environment is recommended to prevent package conflicts.  For stability, a number of dependent packages are installed during the set up of the virtual environment. For \texttt{conda} users the following commands set up the environment (with the initial set of dependencies) and activate it. 
\begin{lstlisting}
$ conda create -n jexosim python=3.8.5 matplotlib=3.3.1-0 setuptools=49.6.0 numpy=1.19.1 numba=0.50.1
$ source activate jexosim
\end{lstlisting}
The activation command might vary with the system.  Next, the Github repository must be cloned.
\begin{lstlisting}
$ git clone https://github.com/subisarkar/JexoSim.git
\end{lstlisting} 
 
\begin{figure}
\begin{center}
    \includegraphics[trim={0 0 0 0}, clip,width=0.7\columnwidth]{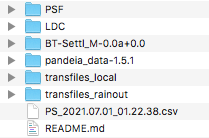}
\caption[]{JexoSim \texttt{archive/} folder displaying the expected contents.}
\label{Fig:archive_folder}
\end{center}
\end{figure}

\noindent Next, the databases described in Section \ref{sec:dependencies} need to be downloaded. These are Pandeia\footnote{https://stsci.app.box.com/v/pandeia-refdata-v1p5p1/}, the Phoenix BT-Settl model grid$^{\ref{note2}}$, and the ATMO transmission spectra grid\footnote{ https://drive.google.com/file/d/ \\1Kvfi7FTBqM1MfnkTnHsJJqI7EFwnuaIG/view}
\footnote{https://drive.google.com/file/d/ \\1Lnp\textunderscore kpbZGPEN0G6QhHrL4DcD4hwvaia/view}  (the latter is optional).
The pre-calculated PSF and LDC folders must also be downloaded and links are provided on the Github `readme' page. The above folders need to be unzipped and moved to inside the \texttt{JexoSim/archive/} directory.  Next, from the NASA Exoplanet Archive$^{\ref{note1}}$, the `Planetery Systems' table (all rows, all columns) should be downloaded in .csv format, and moved into the \texttt{JexoSim/archive} directory.  The final contents of the \texttt{JexoSim/archive/} folder should look similar to Figure \ref{Fig:archive_folder} \footnote{The exact name of the .csv file will vary based on the download date and time but the code will still recognise this as the planetary database. Similarly Pandeia will be recognised even if another version number is used.}.
Next, from within the \texttt{JexoSim/} folder the user should run the following commands to install JexoSim (and remaining dependencies) followed by the generation of JexoSim-compatible transmission files, wavelength solutions, IPC kernels and quantum yields from the Pandeia database.

\begin{lstlisting}
$ cd JexoSim 
$ python setup.py install
$ python generate.py
\end{lstlisting}

\begin{table*}
	\centering
	\caption{Input parameters file options.}
	\label{tab: input parameters}
	\begin{tabular}{p{1.2cm}p{4.2cm}p{3.5cm}p{7.0cm}	}
		\hline		\hline
		Group & Entry & Description & Options\\
		\hline
Simulation 
& \verb'sim_diagnostics' & Display of diagnostic data & 0 or 1\\
& \verb'sim_mode' & Type of simulation recipe & 1: Out-of-transit with Allan analysis. 2: Full transit. 3: Noise budget. 4: Out-of-transit without Allan analysis. \\ 
& \verb'sim_output_type' & Format of the output & 1: Pipeline processed. 2: FITS file only. 3: Pipeline processed with intermediate products.\\
& \verb'sim_noise_source' & Choice of noise source & 	Enter number from provided list corresponding to noise source (e.g. 0 activates all noise sources).\\
& \verb'sim_realisations'&  Number of realisations & Enter integer number of realisations.\\
&  \verb'sim_full_ramps'&  Choice of CDS or UTR &
0: CDS, 1: UTR. \\
& \verb'sim_use_UTR_noise_correction'& Apply noise correction for UTR & If \verb'sim_full_ramps' = 0, choose 0 or 1.\\
& \verb'sim_pointing_psd'& Type of pointing jitter PSD & `flat': white noise PSD, `non-flat': Herschel-derived PSD$^{a}$. \\
&  \verb'sim_prnu_rms'&  rms of PRNU grid (\%) & Enter number. Typical value is 3\%. \\ 
& \verb'sim_flat_field_uncert'& Uncertainty on flat field (\%) &  Enter number. Typical value is 0.5\%. \\
& \verb'sim_use_ipc' & Apply IPC &  0 or 1. Not applied in MIRI.\\
& \verb'sim_use_systematic_model' & Apply systematic model &  0 or 1. .\\
& \verb'sim_systematic_model_name' & Name of the systematic model &  Enter the name of the folder containing the model files\\
\hline
Observation 
& \verb'obs_type'& Observation type  & 1: Primary transit. 2: Secondary eclipse.\\ 
& \verb'obs_frac_T14_pre_transit'& Pre-transit $T_{14}$ fraction & Enter fraction of transit duration pre-transit (e.g. 0.5). \\
& \verb'obs_frac_T14_post_transit'& Post-transit $T_{14}$ fraction & Enter fraction of transit duration post-transit (e.g. 0.5). \\
& \verb'obs_use_sat' & Use saturation time to set $n$ & 0 or 1. \\
& \verb'obs_fw_percent' & Saturation limit & Enter percent of full well (e.g. 80 or 100) as the saturation limit. \\
& \verb'obs_n_groups' & Set number of groups ($n$) & If \verb'obs_use_sat' = 0, enter integer number of groups. \\
& \verb'obs_n_reset_groups' & Set number of reset groups & Enter `default' for default value or integer number of groups$^b$. \\ 
& \verb'obs_inst_config' & Set instrument configuration & Choose from list of configurations with subarray modes.\\
		\hline
Pipeline &  
\verb'pipeline_binning' & Type of spectral binning & `R-bin': uses $R$. `fixed-bin': uses number of pixel columns. \\
& \verb'pipeline_R' & $R$-power for binning & If `R-bin' used enter $R$-power (e.g. 100).\\
& \verb'pipeline_bin_size'	& Pixel columns per bin & If `fixed-bin' used enter number of pixel columns (e.g. 5 or 10). \\ 	
& \verb'pipeline_ap_shape' & Extraction aperture shape	 & `rect': rectangular. `wav': wavelength-dependent. \\ 
& \verb'pipeline_apply_mask' & Apply extraction aperture & 0 or 	 1. \\
& \verb'pipeline_auto_ap' & Auto-select width of aperture & 0 or	  1.	\\
& \verb'pipeline_ap_factor' & Aperture width &   Enter number $\alpha$ where width = 2$\alpha F\lambda$ $^c$.	\\ 	
 & \verb'pipeline_bad_corr'	&  Manage bad pixels & 0 or 1.  \\
		\hline
Exosystem & 
\verb'planet_name' & Planet name &	Enter planet to search for in planet database.\\ 
& \verb'planet_use_database' &	Use planet database & 1: Planet database gives all exosystem parameters. 0: Exosystem parameters entered manually$^d$. \\
& \verb'planet_spectrum_model' & Type of planet spectrum &	`simple': uses flat transmission spectrum based on $(R_p/R_s)^2$ or blackbody-based emission spectrum. `complex': uses ATMO grid$^e$. `file': uses a specific spectrum from a file.\\
& \verb'planet_spectrum_file' & Path to planet spectrum file & If 
\verb'planet_spectrum_model' = `file' provide path to file. \\
 & \verb'star_spectrum_model' & Type of stellar spectrum	& `simple': uses blackbody. `complex': uses Phoenix database. \\
 & \verb'star_spectrum_file' & Path to star spectrum file & If 
\verb'star_spectrum_model' = `file' provide path to file. \\
& \verb'star_spectrum_mag_norm' & Normalise magnitude &    0: None. 1 normalises to J and/or K mag. 2: uses $(R_s/d)^2$\\
	\hline	
		\hline		
\end{tabular}
\begin{tabular}{p{17cm}	}
{\footnotesize$^a$  PSD=power spectral density profile. Both PSDs generate jitter timelines with an rms of 6.7 mas per 1-D axis.  The Herschel-derived PSD is scaled version of the PSD obtained using Herschel pointing information described in \cite{Sarkar2020}.}\\
{\footnotesize$^b$ Default values are 0 for MIRI (assuming a read-reset method), and 1 for all NIR channels.}\\
{\footnotesize$^c$ $F$ is the F-number and $\lambda$ is the wavelength. For rectangular apertures $\lambda$ is the longest wavelength in the band.}\\
{\footnotesize$^d$ There is a section to enter these system parameters manually.}\\
{\footnotesize$^e$ If a `complex' planet spectrum is chosen, a section exists to fill in required parameters (e.g. C/O ratio, haze, cloud etc.)}\\
\end{tabular}	
\end{table*}

\noindent Finally, the user must decide on a directory location for the final results.  By default these will be placed in the \texttt{JexoSim/output/} folder, however the user may choose a different location by editing the `paths file', \texttt{jexosim\textunderscore paths.txt}, which is located in the \texttt{JexoSim/jexosim/input\textunderscore files/} folder. The complete path to the output folder can be entered next to the \texttt{output\textunderscore directory} option.

\subsection{Running a simulation}
\label{sec:running a sim}

To run a simulation the user should navigate to within the \texttt{JexoSim/} folder and run the \texttt{run\textunderscore jexosim.py} file with an input parameters file name (e.g. \texttt{ jexosim\textunderscore input\textunderscore params\textunderscore ex1.txt} as shown below) as the argument. 

\begin{lstlisting}
$ cd JexoSim
$ python run_jexosim.py jexosim_input_params_ex1.txt
\end{lstlisting}

\noindent If the user has chosen to apply the pipeline and obtain processed results,  as mentioned previously these are packaged into a dictionary and dumped in .pickle format in the output directory (Figure \ref{Fig:architecture})  with an associated text file.  The files are identified with the type of simulation, the instrument configuration, the planet, and a time stamp.  To display a set of plots showing results from the given .pickle file, the \texttt{results.py} file is run from within the \texttt{JexoSim/} folder with the filename as an argument.  For example:
\begin{lstlisting}
$ cd JexoSim
$ python results.py Noise_budget_MIRI_LRS_slitless_SLITLESSPRISM_
FAST_HD_209458_b_2021_07_01_1005_28.pickle
\end{lstlisting}

\noindent If the user chooses to simply store the image stack for later analysis without running the pipeline, a FITS file is generated and stored in the output directory (Figure \ref{Fig:architecture}).
 
\subsection{Input parameters file options}

In Table \ref{tab: input parameters} we list the main options in the input parameters file, which provides the interface between the user and the code. These are divided into four categories: `Simulation' (parameters that control the functioning of the simulation, but would not be options for a real observation), `Observation' (parameters that control the simulated observation and could also be options for actual observations), `Pipeline' (parameters controlling the pipeline), and `Exosystem' (parameters for defining the extra-solar system).  As JexoSim evolves, these options may be modified in future.   

\section{Case studies}
We now describe five case studies using JexoSim 2.0.  These demonstrate the range of instrument models (NIRISS, NIRCam, NIRSpec and MIRI), different simulation modes (out-of-transit vs full transit) and different methods to obtain the transit depth precision (Allan deviation analysis vs Monte Carlo).  We use two end-member targets: the hot Jupiter, HD 209458 b, orbiting a bright (J mag 6.59) Sun-like star, and the mini-Neptune, K2-18 b, orbiting an M-dwarf of intermediate brightness (J mag 9.76).
We simulate mostly primary transits, but also consider a secondary eclipse case. Finally, a `noise budget' simulation is performed for MIRI LRS, to show how the total noise can be broken down to show the contribution  of each noise source.

Table  \ref{tab: summary_cases} summarises the major and minor points illustrated by each case study. Each case study can be run by the user by implementing the input parameters files named in Table \ref{tab: summary_cases} and provided with JexoSim 2.0\footnote{These input parameters files are located in the folder \texttt{JexoSim/jexosim/input\textunderscore files/}.} with adjustments described in the text below.

Table \ref{tab: simulation settings} lists the simulation settings for each case study refering to the input parameters file entries given in Table \ref{tab: input parameters}. 

In all cases, except Case 5 (noise budget), the full suite of noise sources, backgrounds and systematics listed in Table \ref{table: summary of noise table} are used (excluding the `user-defined systematic').  For the noise budget case, noise sources are implemented individually as explained further below. 
The JexoSim pipeline (Figure \ref{Fig:architecture}) is used in all cases to obtain the final results.  The results are therefore post-processed, including noise mitigation methods (e.g. aperture masking during 1-D spectra extraction), subtraction of backgrounds, and decorrelation of correlated noise sources (e.g. pointing jitter noise).  


\begin{table*}
	\centering
	\caption{Summary of major and minor points of each case study.}
	\label{tab: summary_cases}
	\begin{tabular}{p{0.3cm}p{1.0cm}p{6.2cm}p{6.2cm}	}
		\hline		\hline
		Case & File & Major points & Minor points\\
\hline
1 
& \verb'*ex1.txt' 
& Produces NIR transmission spectrum of K2-18b
& NIRISS subarray mode selection
\\
& 
& Uses OOT simulation mode 
& Simulation/pipeline settings, e.g.:
\\
{}
& 
& Demonstrates NIRISS instrument model
& auto-calculation of $n$ groups, PRNU level,
\\
&
& Uses Allan deviation method to obtain error bars
& type of jitter PSD, $R$-binning,
\\
&
&  All noise and background sources applied
&  auto-sizing of extraction aperture etc.
\\
&
& Post-processed results via JexoSim pipeline
&  
\\
\hline
2 
& \verb'*ex2.txt' 
& Produces NIR/MIR transmission spectrum of K2-18b
& NIRCam \& MIRI LRS subarray mode selection
\\
& 
& Uses full-transit simulation mode 
& Simulation/pipeline settings, e.g.:
\\
& 
& Demonstrates NIRCam \& MIRI instrument models
& pre- and post- transit observation time
\\
&
& Uses Monte Carlo method to obtain error bars 
& Combines spectra from several instrument modes
\\
&
& All noise and background sources applied
& 
\\
&
& Post-processed results via JexoSim pipeline
& 
\\
\hline
3
& \verb'*ex3.txt' 
& Produces NIR transmission spectrum of K2-18b
& NIRSpec subarray mode selection
\\
& 
& Produces NIR transmission spectrum of HD 209458 b
& Simulation/pipeline settings, e.g.:
\\
& 
& Uses full-transit simulation mode
& number of realisations set to 50
\\
&
& Demonstrates NIRSpec instrument model	
& Combines spectra from several subarray modes
\\
&
& Uses Monte Carlo method to obtain error bars
& 
\\
&
& All noise and background sources applied
&  
\\
&
& Post-processed results via JexoSim pipeline
& 
\\
&
& Saturation affecting subarray choices
&
\\
&
& Comparison of hot Jupiter / sub-Neptune spectrum SNR
&
\\
\hline
4
& \verb'*ex4.txt' 
& Produces NIR/MIR emission spectrum of HD 209458 b 
& Uses same subarray modes as in case 2
\\
&
&  Uses full-eclipse simulation mode
&  Simulation/pipeline settings, e.g.
\\
& 
& Demonstrates NIRCam and MIRI instrument models
& selection of eclipse over transit observation
\\
& 
& Uses Monte Carlo method to obtain error bars
& Combines spectra from several instrument modes
\\
& 
& All noise and background sources applied
&
\\
& 
& Post-processed results via JexoSim pipeline
&
\\
\hline
5
& \verb'*ex5.txt' 
& Produces noise budget for MIRI K2-18b observation 
& Uses LRS SLITLESSPRISM subarray mode
\\
& 
& Uses Allan deviation method to find fractional noise
& Simulation of multiple noise sources in isolation
\\
& 
& Significant background \& read noise at longer $\lambda$
& Simulation/pipeline settings, e.g.
\\
& 
& 
& user-defined aperture size (auto-sizing disabled)
\\
	\hline	
		\hline		
\end{tabular}
\begin{tabular}{p{17cm}	}
{\footnotesize$^*$ Prefix: \texttt{jexosim\textunderscore input\textunderscore params\textunderscore}}\\
\\
\end{tabular}	
\end{table*}


\begin{table*}
	\centering
	\caption{Simulation settings per case study.  See Table \ref{tab: input parameters} for description of each entry.  Some entries are left blank indicating they do not need to be filled in for these simulations.}
	\label{tab: simulation settings}
	\begin{tabular}{p{0.5cm}p{3.8cm}ccccc	}
		\hline		
		\hline
		Group & Entry & Case 1 & Case 2 & Case 3 & Case 4 & Case 5\\
		\hline
 \parbox[t]{2mm}{\multirow{11}{*}{\rotatebox[origin=r]{90}{Simulation      }}}
& \verb'sim_diagnostics' & 0 & 0 & 0 & 0 & 0\\
& \verb'sim_mode' & 1 & 2 & 2 & 2 & 3\\ 
& \verb'sim_output_type' & 1 & 1 & 1 & 1 & 1\\ 
& \verb'sim_noise_source' & 0 & 0 & 0 & 0 &  \\
& \verb'sim_realisations'& 5 & 50 & 50 & 50 & 10 \\ 
&  \verb'sim_full_ramps'& 0 & 0 & 0 & 0 & 0 \\
& \verb'sim_use_UTR_noise_correction'& 1 & 1 & 1 & 1 & 1 \\
& \verb'sim_pointing_psd'& `flat' & `flat' & `flat' & `flat' & `flat' \\
&  \verb'sim_prnu_rms'& 3 & 3 & 3 & 3 & 3 \\
& \verb'sim_flat_field_uncert'& 0.5 & 0.5 & 0.5 & 0.5 & 0.5 \\
& \verb'sim_use_ipc'  & 1 & 1 & 1 & 1 & 1 \\
& \verb'sim_use_systematic_model' & 0 & 0 & 0 & 0 & 0 \\
& \verb'sim_systematic_model_name' &   &   &   &   &  \\

\hline
 \parbox[t]{2mm}{\multirow{12}{*}{\rotatebox[origin=r]{90}{Observation      }}}
& \verb'obs_type'& 1 & 1 & 1 & 2 & 1 \\
& \verb'obs_frac_T14_pre_transit' &  & 0.5 & 0.5 & 0.5 &  \\
& \verb'obs_frac_T14_post_transit'&  & 0.5 & 0.5 & 0.5 &  \\
& \verb'obs_use_sat' & 1 & 1 & 1 & 1 & 1 \\
& \verb'obs_fw_percent' & 80 & 80 & 80 & 80 & 80 \\
& \verb'obs_n_groups' &   &   &  &   &   \\
& \verb'obs_n_reset_groups' & `default' & `default' & `default' & `default' & `default' \\
& \verb'obs_inst_config'$^a$ & `NIRISS + & `NIRCam + &  `NIRSpec &  `MIRI + & `MIRI +   \\
&   &  SOSS\textunderscore  & TSGRISM\textunderscore &  BOTS\textunderscore G395H\textunderscore & LRS\textunderscore & LRS\textunderscore    \\
&   &  GR700XD + & F322W2 + & F290LP +   & slitless + &   slitless + \\
&   & SUBSTRIP96 +  & SUBGRISM64\textunderscore  &  SUB2048 +  & SLITLESSPRISM +  & SLITLESSPRISM +\\
&  & NISRAPID'  & 4\textunderscore output + &  NRSRAPID' &   FAST'  &    FAST'\\
&  & &  RAPID' &   &   &    \\

		\hline
		
 \parbox[t]{2mm}{\multirow{7}{*}{\rotatebox[origin=r]{90}{Pipeline     }}} &  
\verb'pipeline_binning' & `R-bin' & `R-bin'  & `R-bin'  & `R-bin' & `R-bin'    \\
& \verb'pipeline_R' & 100 & 100  & 100  & 100 & 100    \\
& \verb'pipeline_bin_size'	&   &    &   &   &  \\ 	
& \verb'pipeline_ap_shape' &  `rect' &  `rect'  & `rect'  &  `rect' &  `rect' \\ 	
& \verb'pipeline_apply_mask' &  1 &  1  &  1 &  1 &  1\\ 
& \verb'pipeline_auto_ap' &  1 &  1  &  1 &  1 &  0\\ 
& \verb'pipeline_ap_factor' &   &     &    &    &  1\\ 	
 & \verb'pipeline_bad_corr'	&  1   &  1   &   1 &  1  &  1\\ 	
		\hline
		
 \parbox[t]{2mm}{\multirow{6}{*}{\rotatebox[origin=r]{90}{Exosystem      }}} & 
\verb'planet_name'$^b$ & `K2-18 b'  & `K2-18 b'   & `HD 209458 b'    & `HD 209458 b'   &  `K2-18 b'\\ 	
& \verb'planet_use_database' &  1   &  1   &   1 &  1  &  1\\ 
& \verb'planet_spectrum_model' &   `file'   &  `file'   &   `file' &  `file' &  `file'\\ 
& \verb'planet_spectrum_file' &  path given   &  path given   &   path given &  path given  &  path given\\
 & \verb'star_spectrum_model' &   `complex'   &  `complex'   &   `complex' &  `complex' & `complex'\\ 
 & \verb'star_spectrum_file' &     &    &   &   &  \\
& \verb'star_spectrum_mag_norm' & 1   &  1   &  1 &  1  &  1\\
	\hline	
		\hline		
\end{tabular}
\begin{tabular}{p{17cm}	}
{\footnotesize$^a$  For cases 2-4, this entry is changed as required for the instrument configuration being simulated.}\\
{\footnotesize$^b$ In case 3 `HD 209458 b' is also used.}\\
\end{tabular}	
\end{table*}


\subsection{Case 1: Transmission spectrum of K2-18 b with NIRISS using Allan deviation method}
\subsubsection{Overview}
For this case study we simulate NIRISS SOSS (single object slitless spectroscopy) with the GR700XD grism and the SUBSTRIP96 subarray in NISRAPID read out mode.  K2-18 b, is selected as the target, with all parameters obtained via the planet database.  The goal is to obtain the precision on the transit depth at a binned $R$-power of 100 and produce example transmission spectra with error bars.  We utilise an out-of-transit simulation approach (i.e. no transit light curves implemented) and Allan deviation analysis of the results to obtain the precision. The advantages and disadvantages of this approach are discussed below.

\subsubsection{Allan deviation analysis}
Allan deviation analysis \citep{Allan1966} accounts for the effect of correlated noise on the standard error of a mean signal. This is achieved by taking a time-dependent signal and binning it into sequential segments of duration $\tau$, then taking the mean in each segment, and measuring the standard error of the mean values. For uncorrelated (`white') noise, the fractional noise (the standard error of the mean / mean signal), falls predictably with $\tau$ with a power law exponent of -0.5. This is the basis of `integrating down' photon noise. However for correlated noise, the exponent may be different and itself change with $\tau$.  At large enough values of $\tau$ the correlation timescale is exceeded and the noise integrates down like `white' noise.  In \cite{Sarkar2020} we described in detail how JexoSim performs this kind of analysis using simulated out-of-transit spectrally-binned timelines of integration signals.  The outcome of this analysis is that the fractional noise, $\sigma_s$, can be predicted for any value of $\tau$, taking into account any correlated noise.  If $\tau$ is set to $T_{14}$, the transit duration, then assuming a simple `box-car' model of a transit, with an equal amount of out-of-transit and in-transit time, the noise on the transit depth can be approximated by $\sqrt{2}\sigma_s$.

This approach partially accounts for correlated noise (by finding the standard error on the mean in- and out-of-transit signals per spectral bin), but assumes the in- and out-of-transit light curve sections are uncorrelated to each other. It also does not account for limb-darkening on the light curve, and cannot be used to assess systematic biases distorting the light curve, such as spot occultations or persistence.   Full transit simulations using the Monte Carlo approach described in the next case study can capture these effects.  However for the assessment of transit depth precision in the absence of such light curve distortions, this approach is adequate and delivers equivalent results to the Monte Carlo approach.

\subsubsection{Simulation settings}
Key settings in the input parameters file for this case are shown in Table \ref{tab: simulation settings}.  The option \texttt{sim\textunderscore mode} is set to 1 choosing an OOT simulation with Allan analysis and \texttt{sim\textunderscore output\textunderscore type} set to 1, selecting pipeline processed results.
The jitter option \texttt{sim\textunderscore pointing\textunderscore psd} is set to `flat' selecting a white noise power spectrum generating a jitter timeline with rms 6.7 mas. The PRNU is set to 3\% rms with a flat field uncertainty of 0.5\%.
The number of groups per integration is set by the saturation time which is the time taken for the first pixel on the detector to reach the saturation limit.  This time is then divided by the group time (same as frame time when $m$ = 1) to obtain the integer, $n$, the number of groups per integration.  80\% of the pixel full-well is chosen as the saturation limit. In this case $n$ was calculated to be 13.  For speed we implement CDS rather than full ramps but apply the UTR noise correction that adjusts the Poission and read noise to the levels expected for the number of groups per integration.

The option \texttt{planet\textunderscore spectrum\textunderscore model}  is set to `file' so that the transmission spectrum is obtained from a specific file generated beforehand for this planet using a radiative transfer code. In our examples, we use models based on published works for each planet considered. We use a model transmission spectrum of HD~209458~b based on \cite{Welbanks2019} and that of K2-18~b based on \cite{Madhusudhan_2020_K2}. The pipeline option, \texttt{pipeline\textunderscore binning} is set to `R-bin' for spectral binning based on spectral resolving power, with the $R$-power set to 100.  \texttt{pipeline\textunderscore ap\textunderscore shape} is set to `rect' choosing a rectangular extraction aperture.
Since in NIRISS SOSS the spectral trace is curved, choosing this latter option does not actually produce a rectangular-shaped aperture. Rather the aperture is curved to match the shape of the spectrum itself but the width of the aperture remains constant with wavelength. By setting \texttt{pipeline\textunderscore auto\textunderscore ap} to 1, the width of the aperture is auto-selected by the code to optimise SNR.  We use the average of 5 realisations to obtain the final noise results, by setting \texttt{sim\textunderscore realisations} to 5\footnote{One realisation would suffice here, but obtaining the average of many gives a less noisy final result.}.

\subsubsection{Results}

\begin{figure*}
\begin{center}
\captionsetup{justification=centering}
  \begin{subfigure}[b]{0.6\textwidth}
    \includegraphics[trim={-2cm 1cm 0 0}, clip,width=\textwidth]{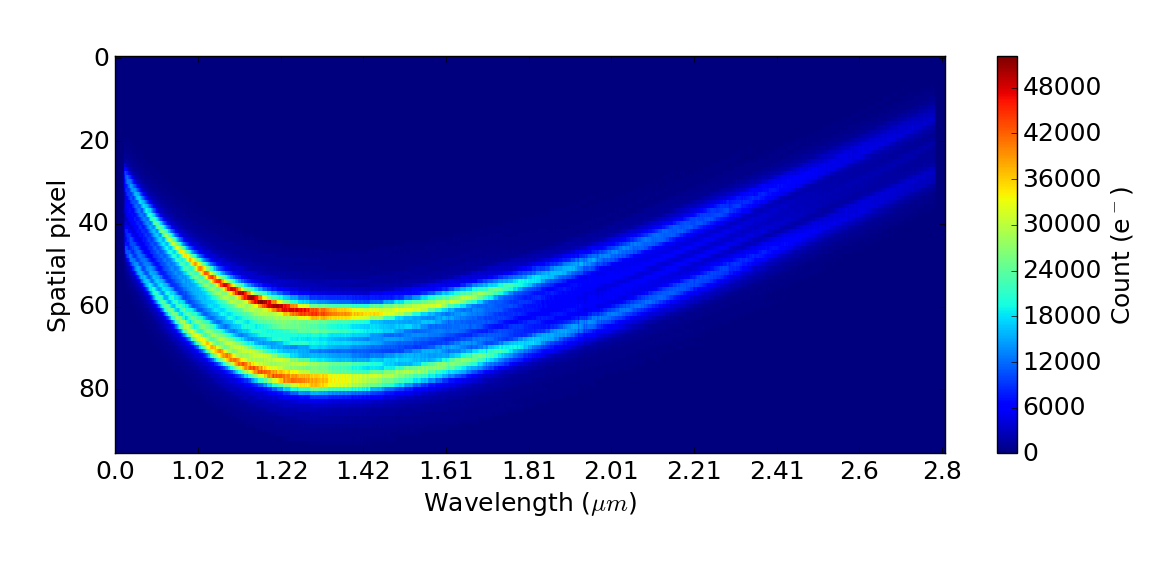}
    \caption{ }
    \label{fig:1}
  \end{subfigure}
  \begin{subfigure}[b]{0.45\textwidth}
    \includegraphics[trim={0 0 0 0}, clip,width=\textwidth]{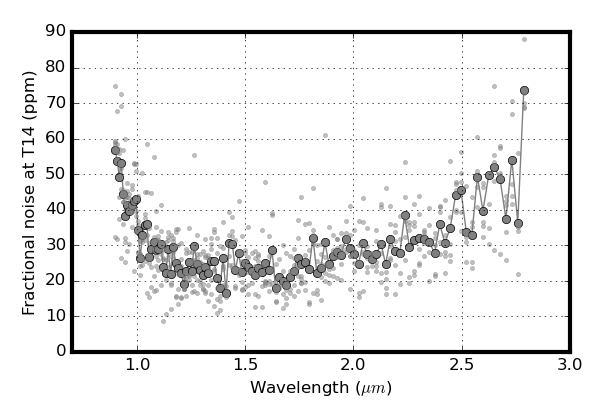}
    \caption{ }
    \label{fig:1}
  \end{subfigure}
  \begin{subfigure}[b]{0.45\textwidth}
    \includegraphics[trim={0 0 0 0}, clip,width=\textwidth]{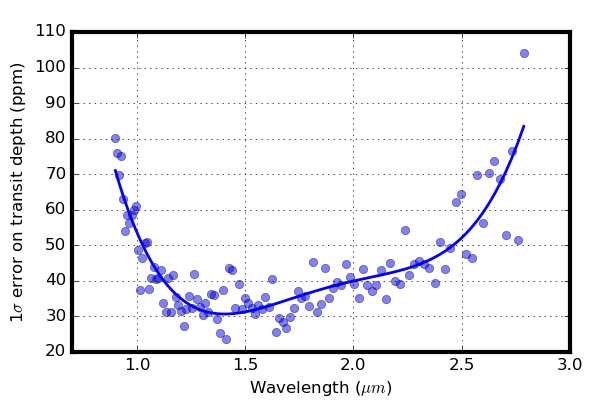}
    \caption{ }
    \label{fig:1}
  \end{subfigure}
  \begin{subfigure}[b]{0.80\textwidth}
    \includegraphics[trim={0 0 0 0cm}, clip,width=\textwidth]{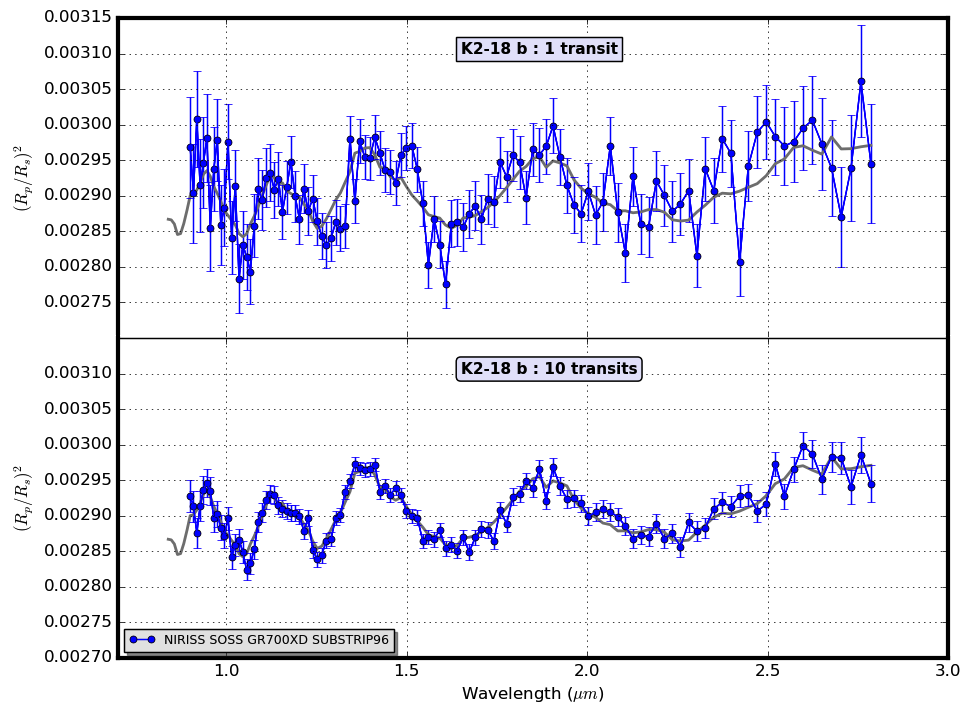}
    \caption{ }
    \label{fig:1}
  \end{subfigure}
\caption[]{Case 1: simulation of transmission spectrum for K2-18 b with NIRISS SOSS GR700XD SUBSTRIP96 using the Allan deviation method. a) Integration image from pipeline. b) Fractional noise at $T_{14}$ versus wavelength. Small dots: individual realisations. Large dots: average result.  c) 1$\sigma$ error on transit depth versus wavelength with 4th order polynomial line fit. d) Synthetic post-processed transmission spectra at $R$ = 100 for 1 transit and 10 co-added transits. Grey line: input spectrum binned to $R$ = 100.}
\label{Fig:ex1}
\end{center}
\end{figure*}

Figure \ref{Fig:ex1} shows the results from this case study.  Figure \ref{Fig:ex1}a shows an example  integration image taken just before the jitter decorrelation step in the pipeline (Figure \ref{Fig:architecture}). We see the curved spectral trace in this channel and the unusual PSF, elongated in the cross-dispersion direction by a cylindrical lens incorporated into the GR700XD optical element.
Figure \ref{Fig:ex1}b shows the fractional noise at $T_{14}$ as a function of wavelength derived from the Allan deviation analysis. The average result is given by the large grey dots, with the small dots showing the individual realisations. 
The average value is used to then calculate the noise on the transit depth per spectral bin (Figure \ref{Fig:ex1}c, blue dots) by multiplying by $\sqrt{2}$. We fit a 4th order polynomial to these points (Figure \ref{Fig:ex1}c, blue line) to estimate the expected noise (1$\sigma$ error) per spectral bin. 

We find that at $R$ = 100, the 1$\sigma$ error given by the polynomial fit ranges from a minimum of 31 ppm (at 1.41 \textmu m) to maximum of 83 ppm (at 2.79 \textmu m), the noise rising at both the blue and red ends of the wavelength range.  We then generate a transmission spectrum in Figure \ref{Fig:ex1}d.  In each spectral bin, the 1$\sigma$ error from \ref{Fig:ex1}b line fit is used to randomise the value of $(R_p/R_s)^2$ using a Gaussian probability distribution centered on the input spectrum binned down to $R$ = 100.  The input spectrum is the planet spectrum modulated by the instrument response: $p(X)$ in Figure \ref{Fig: algo}.  The size of the error bars represent the 1$\sigma$ margin of error.  The upper plot in Figure \ref{Fig:ex1}d gives a randomised spectrum for a single transit observation.  

The spectral features are noisy but large scale variations are discernible over the noise, including the 1.4 \textmu m water featured discovered using the Hubble WFC3 \citep{Benneke_2019, Tsiaris2019}.  A future spectral retrieval study can quantify observability more exactly. The lower plot shows a spectrum for 10 co-added transits where the 1$\sigma$ error has been reduced by a factor of $1/\sqrt{10}$. We can see that with 10 transits the spectral features are well-recovered with very little noise at the shorter wavelengths.

\subsection{Case 2: Transmission spectrum of K2-18 b with NIRCam and MIRI using Monte Carlo method}
\label{sec:Case 2: K2-18 b with NIRCam and MIRI using Monte Carlo method}

\subsubsection{Overview}
In this case study we simulate the NIRCam Time-series Grism F444W and F322W2 modes observing K2-18 b in primary transit.  In both cases, the SUBGRISM64 subarray is used with the 4-output readout pattern option.
We also simulate MIRI LRS with the SLITLESSPRISM subarray.  For each instrument mode, full primary transit observations are modeled in a 50-realisation Monte Carlo simulation to obtain the precision on the transit depth.  The pipeline runs through stage 1 and stage 2 (light curve fitting) (Figure \ref{Fig:architecture}).  This combination of instrument modes allows us to generate a simulated transmission spectrum from 2.4 to 12 \textmu m.

\subsubsection{Monte Carlo method}
This method generates full primary transit or secondary eclipse simulations with implementation of a transit or eclipse light curve.  In each realisation, the image time-series are processed through stage 1 of the pipeline generating synthetic data light curves spectrally-binned to $R$ of 100.  In stage 2 of the pipeline, these are fitted with model light curves and the transit depth, $(R_p/R_s)^2$ or eclipse depth, $F_p/(F_s+F_p)$, recovered for each spectral bin.  The fitting  uses a downhill simplex algorithm minimising chi-squared. The planet-star radius ratio (or square root of the eclipse depth in secondary eclipse simulations) is the only free-parameter in this fit. After multiple realisations, a distribution of transit depths is obtained for each spectral bin.  We take the standard deviation of this distribution to be the 1$\sigma$ confidence interval margin-of-error on the transit depth, i.e. the size of the error bar on the final spectrum.  The Monte Carlo approach gives robust results since the uncertainty on the transit depth is directly measured, accounting for all correlated and uncorrelated noise effects.  Also, any residual systematic wavelength-dependent biases on the spectrum can be detected through deviation of the mean value in a systematic manner relative to the input spectrum. The latter, for example, may occur due to the presence of stars spots and faculae.

\subsubsection{Simulation settings}
Table \ref{tab: simulation settings} gives the key settings used in the input parameters file for this case study.
These are mostly the same as used for case 1, with the following exceptions.  
\texttt{sim\textunderscore mode} is set to 2, choosing a full transit observation, and \texttt{sim\textunderscore realisations} is set to 50.  Under 
\texttt{obs\textunderscore inst\textunderscore config}, the option for NIRCam F332W2 SUBGRISM64 with 4-output readout pattern is shown, however this is substituted when simulating the F444W and MIRI LRS modes.  The fraction of $T_{14}$ that is observed before transit (\texttt{obs\textunderscore frac\textunderscore T14\textunderscore pre\textunderscore transit})  and the fraction after transit (\texttt{obs\textunderscore frac\textunderscore T14\textunderscore post\textunderscore transit}) are both set to 0.5.

\subsubsection{Results}

\begin{figure*}
\begin{center}
\captionsetup{justification=centering}
  \begin{subfigure}[b]{1.0\textwidth}
    \includegraphics[trim={0cm 0 0 0}, clip,width=0.3\textwidth]{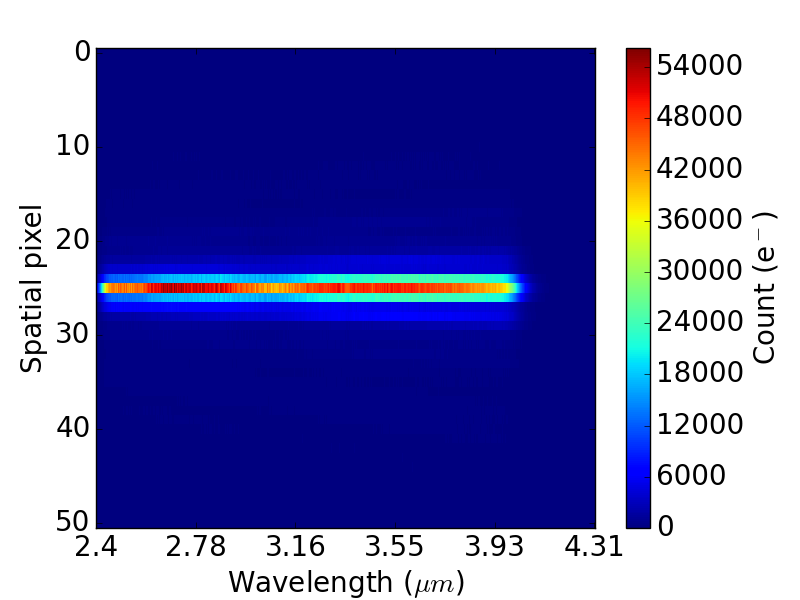}
    \includegraphics[trim={0cm 0 0 0}, clip,width=0.3\textwidth]{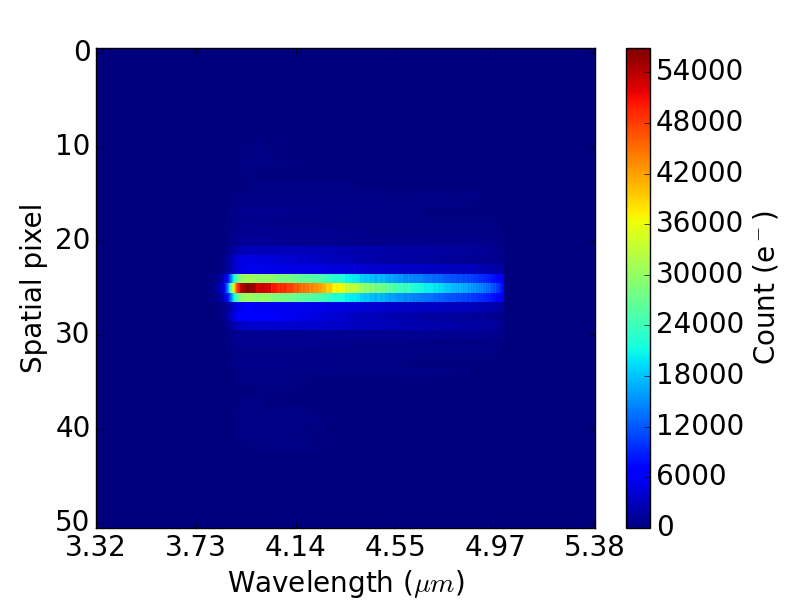}
    \includegraphics[trim={0cm 0 0 0}, clip,width=0.3\textwidth]{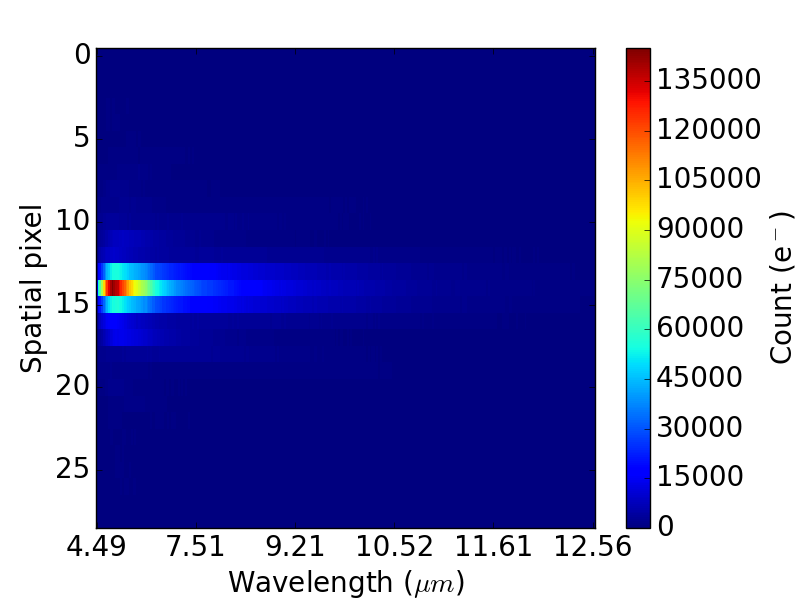} 
    \caption{ }
    \label{fig:1}
  \end{subfigure}
  \begin{subfigure}[b]{0.45\textwidth}
    \includegraphics[trim={0 0 0 0}, clip,width=\textwidth]{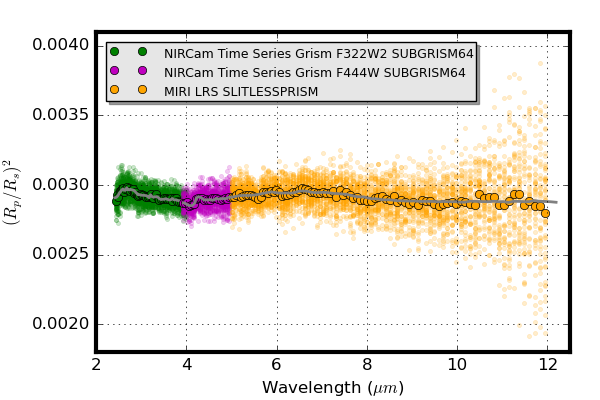}
    \caption{ }
    \label{fig:1}
  \end{subfigure}
  \begin{subfigure}[b]{0.45\textwidth}
    \includegraphics[trim={0 0 0 0}, clip,width=\textwidth]{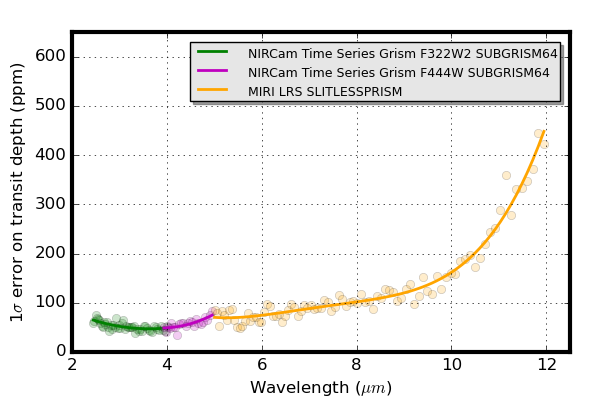}
    \caption{ }
    \label{fig:1}
  \end{subfigure}
  \begin{subfigure}[b]{0.80\textwidth}
    \includegraphics[trim={0 0 0 0cm}, clip,width=\textwidth]{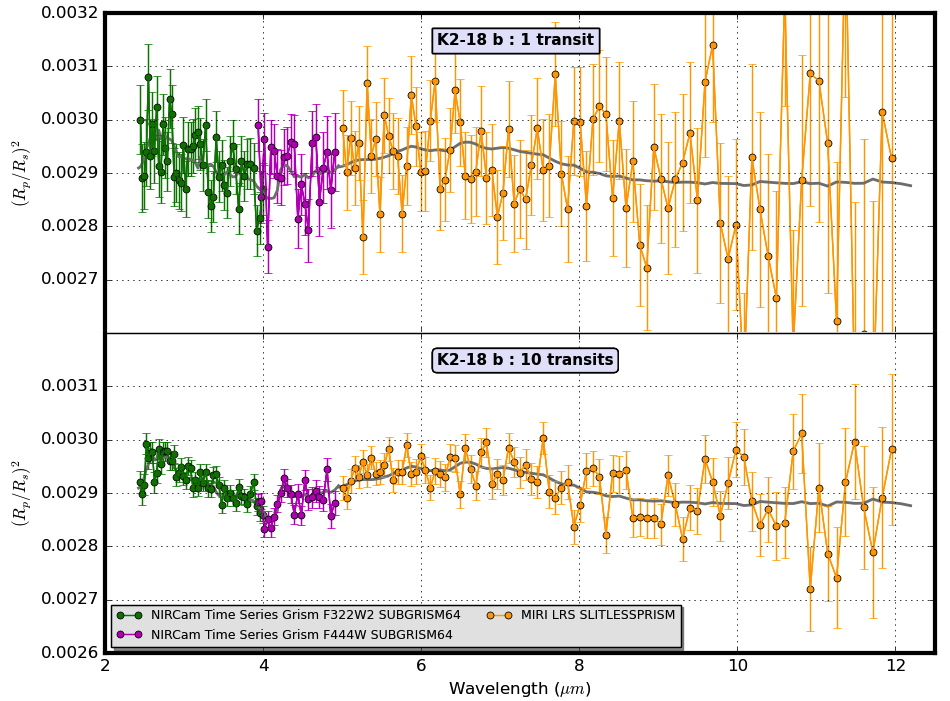}
    \caption{ }
    \label{fig:1}
  \end{subfigure}
\caption[]{Case 2: simulation of transmission spectrum for K2-18 b with NIRCam Time Series Grism F322W2 and F444W modes (both with SUBGRISM64) and MIRI LRS slitless using the Monte Carlo method.  a) Integration images from pipeline - left: NIRCam Grism F322W2, middle: NIRCam Grism F444W, right: MIRI LRS. b) Monte Carlo simulation recovered transit depths versus wavelength (large dots: average of 50 realisations, small dots: individual realisations, grey line:  input spectrum binned to $R$ = 100).
 c) 1$\sigma$ error on transit depth versus wavelength with 2nd and 4th order polynomial line fits for NIRCam and MIRI respectively.
d) Synthetic post-processed transmission spectra for K2-18b for 1 transit and 10 co-added transits at $R$= 100  (grey line: input spectrum binned to $R$ = 100)}
\label{Fig:ex2}
\end{center}
\end{figure*}

Results from this case study are shown in Figure \ref{Fig:ex2}.  Example integration images are shown for each instrument configuration in Figure \ref{Fig:ex2}a. Due to cropping of the image in the `Noise' module to improve simulation speed, the size of these images may not match the original subarray sizes.
Figure \ref{Fig:ex2}b shows the transit depths recovered from each of the 50 realisations (smaller dots). In each bin a distribution of transit depth instances occurs. The mean value for the distribution in each spectral bin is shown (large dots). The input spectrum binned to $R$ = 100 is shown as the grey line.

No discernible systematic bias is seen in the mean values.  Figure \ref{Fig:ex2}c shows the 1$\sigma$ error versus wavelength, where 
each point is the standard deviation of transit depths from Figure \ref{Fig:ex2}b  for the given spectral bin. Polynomial fits to these points (4th order for MIRI and 2nd order for NIRCam) are shown. These line fits are used to predict the 1$\sigma$ error for each wavelength bin for a single transit observation per instrument mode.

Based on these polynomial fits, the 1$\sigma$ error in NIRCam ranges from 47 ppm (at 3.61 \textmu m) to 65 ppm (at 2.45 \textmu m) for the F332W2 mode, and from 49 ppm (at 3.94 \textmu m) to 75 ppm (at 4.95 \textmu m) for the F444W mode.  Similarly, for MIRI LRS, the 1$\sigma$ error ranges from a minimum of 70 ppm (at 5.26 \textmu m) increasing substantially with wavelength beyond 10 \textmu m to a maximum of 448 ppm (at 11.95 \textmu m).

As in the previous case study we generate example transmission spectra using these 1$\sigma$ error values. Figure \ref{Fig:ex2}d shows randomised transmission spectra for 1 and 10 transits.  With one transit the spectrum is very noisy with features not readily discernible over noise. With 10 transits, the spectral features in both NIRCam and MIRI become discernible above noise upto about 8 \textmu m range.

\subsection{Case 3: Transmission spectra of HD 209458 b and K2-18 b with NIRSpec using Monte Carlo method}
\label{sec:Case 3: HD 209458 b and K2-18 b with NIRSpec using Monte Carlo method}

\subsubsection{Overview}
This case study simulates NIRSpec with a variety of subarray modes, using both medium- and high-resolution gratings.  Simulated primary transit observations are performed on both HD 209458 b and K2-18 b.  This example also illustrates the potential issues of detector saturation for bright targets with JWST requiring careful subarray selection, with different configurations chosen for each planet.
We obtain the transit depth precision at $R$ = 100 for each configuration and planet using the Monte Carlo method described in Section \ref{sec:Case 2: K2-18 b with NIRCam and MIRI using Monte Carlo method}. Using a combination of instrument configurations we generate post-processed transmission spectra with error bars for each planet from $\sim$ 1-5 \textmu m. These results highlight the comparative performance of JWST NIRSpec between a hot Jupiter with a bright host star and a mini-Neptune with a dim host star.

\subsubsection{Simulation settings}
Table \ref{tab: simulation settings} lists the input parameters file settings for this example.  These mostly follow those of case 2.  The entry for \texttt{obs\textunderscore inst\textunderscore config} is shown for the G395H grating with the F290LP filter and the SUB2048 subarray, but this is changed for other configurations as needed. The  \texttt{planet\textunderscore name} entry is changed between `HD 209458 b' and `K2-18 b'.

\subsubsection{NIRSpec configurations}
 NIRSpec is a complex instrument with several different dispersion elements, filter options and subarray modes  (Figure \ref{Fig:PCE}). The choice of instrument configuration (grating, filter and subarray) may be driven by the desired wavelength coverage and the wish to minimise the number of separate transit observations needed to cover this wavelength range. Each configuration requires an independent observation.

 However configuration choice must also take into account the feasibility of using each subarray with regards to pixel saturation.  In particular `hard' saturation must be avoided.  This is where the saturation time (the time for the first pixel to reach the chosen saturation limit) is longer than 2 frame times.  2 frames is the minimum needed to mitigate reset noise. Subarrays with shorter frame times are thus more likely to avoid hard saturation. As with the previous simulations, we use a saturation limit of 80\% full well (to minimise potential for non-linearity) and assume the NRSRAPID readout mode.

For both HD 209458 b and K2-18 b, JexoSim predicts NIRSpec PRISM will result in saturation with all subarray modes and is thus excluded in these cases. This is partly attributable to its very short spectral trace, concentrating spectral power on a small region of the detector, leading to high detector count rates and short saturation times.

The SUB512 and SUB512S subarrays have the shortest frame times and are thus least prone to saturation.  However, as can be seen in Figure \ref{Fig:PCE} they cover only a fraction of the available wavelength range for a given dispersion element and so would need to be combined with other subarrays for wide wavelength coverage.

The high resolution gratings, which tend to disperse the spectral power over a greater physical distance compared to the medium resolution gratings, are better for avoiding saturation.  For the brighter target, HD 209458 b, we found we could apply G395H using the SUB2048 subarray (thus capturing its full wavelength range) without saturation. For G295H, we found that SUB2048 leads to hard saturation, but subarrays SUB1024A and SUB1024B (which have shorter frame times) avoid this, and thus a combination of these two subarray modes is used.  Finally for G140H, we found that subarrays SUB1024A and SUB1024B saturate, but subarray SUB512 is feasible with the F100LP filter giving coverage in two wavelength patches (1.07–1.19, 1.47–1.59 \textmu m) (Figure \ref{Fig:PCE}).

For K2-18 b, we found that all three medium resolution gratings (G140M, G235M, and G395M) could be used with the SUB2048 subarray (capturing the full wavelength range of each grating) without saturating.  We chose the F100LP filter for use with G140M over the the alternate F070LP filter
as it avoids a gap in the wavelength coverage when combined with the other gratings (Figure \ref{Fig:PCE}).

\subsubsection{Results}

\begin{figure*}
\begin{center}
\captionsetup{justification=centering}
  \begin{subfigure}[b]{0.45\textwidth}
    \includegraphics[trim={0 0 0 0}, clip,width=\textwidth]{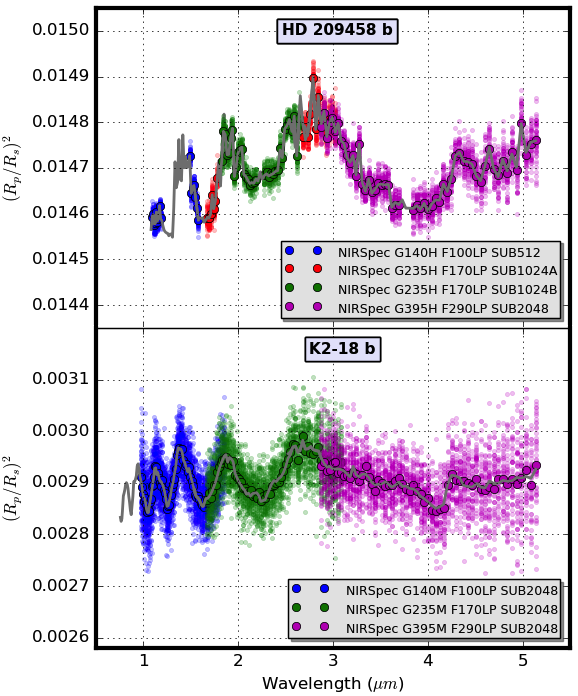}
    \caption{ }
    \label{fig:1}
  \end{subfigure}
  \begin{subfigure}[b]{0.42\textwidth}
    \includegraphics[trim={0 0 0 0}, clip,width=\textwidth]{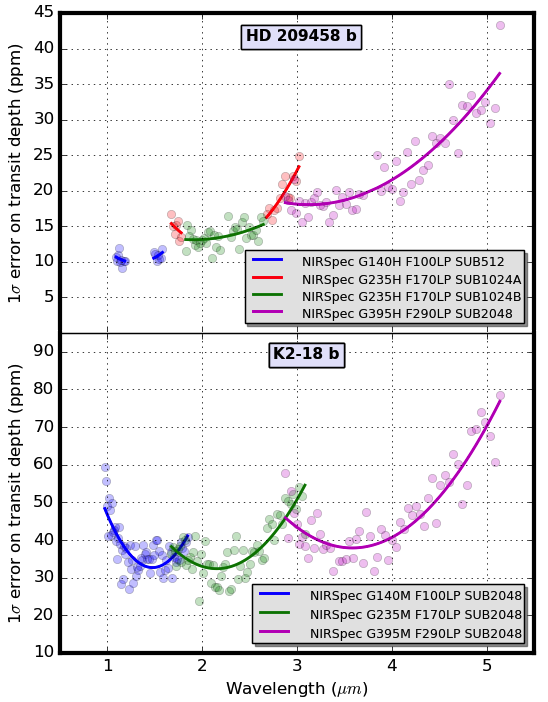}
    \caption{ }
    \label{fig:1}
  \end{subfigure}
  \begin{subfigure}[b]{0.80\textwidth}
    \includegraphics[trim={0 0 0 0cm}, clip,width=\textwidth]{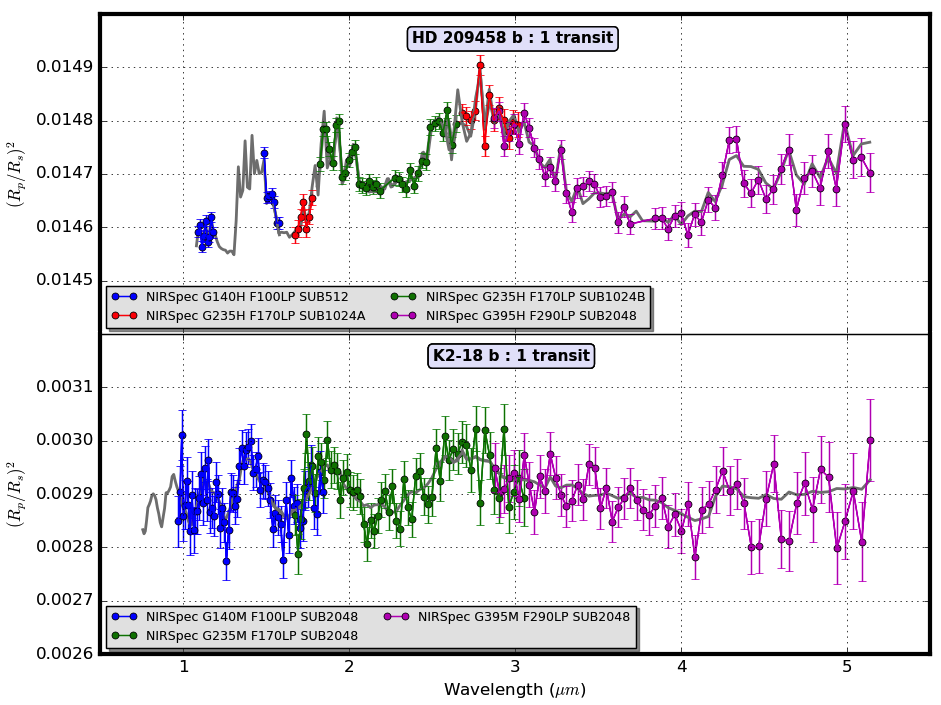}
    \caption{ }
    \label{fig:1}
  \end{subfigure}
\caption[]{Case 3: simulation of transmission spectra for HD209458b and K2-18b using multiple NIRSpec instrument subarray modes using the Monte Carlo method.
a) Monte Carlo simulation recovered transit depths versus wavelength (large dots: average of 50 realisations, small dots: individual realisations, grey line:  input spectrum binned to $R$ = 100).
 b) 1$\sigma$ error on transit depth versus wavelength with 2nd order polynomial line fits.
c) Synthetic post-processed transmission spectra for each planet at $R$= 100 (grey line: input spectrum binned to $R$ = 100).}
\label{Fig:Case3}
\end{center}
\end{figure*}

Figure \ref{Fig:Case3} show the results for this case study.  The distributions and means of transit depth instances recovered from the Monte Carlo simulation are shown in Figure \ref{Fig:Case3}a.  No discernible biases are seen when comparing the mean transit depths to the input spectrum.  The standard deviations of the transit depths in each spectral bin are plotted in Figure \ref{Fig:Case3}b, with second order polynomials fitted to the values.

Using the polynomial values, for HD 209458 b the 1$\sigma$ error on the transit depth across all the NIRSpec modes used ranges from 10 ppm (at 1.18 \textmu m) to 36 ppm (at 5.14 \textmu m).  For K2-18 b the 1$\sigma$ error across the NIRSpec modes used ranges from 32 ppm (at 2.15 \textmu m) to 77 ppm (at 5.14 \textmu m).  The noise on the transit depth is thus about 2-3 times higher for K2-18 b compared to HD 209458 b.

In additional to experiencing higher noise, the spectral amplitudes for the low-mass K2-18 b are much smaller than for the hot Jupiter HD 209458 b (Figure \ref{Fig:Case3}c) which further reduces the final signal-to-noise on the spectrum (where the `signal' is the spectral amplitude) compared to HD 209458 b.
We can readily see from Figure \ref{Fig:Case3}c that the transmission spectrum for HD 209458 b recovers planetary spectral features to a high degree of precision and accuracy.  For K2-18 b, large NIR spectral features (such as the 1.4 \textmu m water feature) are discernible above noise.  The final spectrum for HD209458b contains a gap in wavelength coverage between 1.19 and 1.47 \textmu m due to using the SUB512 subarray for the G140H grating.

\subsection{Case 4: Day-side emission spectrum of HD 209458 b with NIRCam and MIRI using Monte Carlo method}
 
\subsubsection{Overview}
This case illustrates a simulated observation of the secondary eclipse of HD 209458 b using NIRCam and MIRI LRS.  The same instrument configurations used in case 2 are used here (NIRCam Time-series Grism F444W and F332W2 with 4-output readout pattern and SUBGRISM64 subarray and MIRI LRS SLITLESSPRISM).   The simulations use eclipse light curves  to modulate the stellar signal.  These are generated in the same way as primary transit light curves but limb-darkening coefficients are set to zero and the minimum of the light curve is normalised to unity.  The fractional eclipse depth of each light curve gives the planet-star flux ratio expressed as $F_p/(F_s+F_p)(\lambda)$.
The input planet spectrum for HD~209458~b was obtained from a self-consistent model reported in \cite{Gandhi2017}.  As in the transmission cases, all noise sources, background sources and instrumental effects in Table \ref{table: summary of noise table} (excluding the user-defined systematic) are included, and the final results are processed through the pipeline in the same way as for cases 2 and 3.  The Monte Carlo approach is used with 50 realisations for each instrumental configuration.  We find the precision on the eclipse depth for an $R$ of 100 and produce a post-processed day-side emission spectrum extending from 2.4-12 \textmu m.

\subsubsection{Simulation settings}
The settings for this simulation are given in Table \ref{tab: simulation settings} and are mostly the same as used in case 2 .  The key change is for \texttt{obs\textunderscore type} where 2 is selected to generate a secondary eclipse observation.  `HD 209458 b' is entered for \texttt{planet\textunderscore name}. The path to the secondary eclipse input planet spectrum file is entered under  \texttt{planet\textunderscore spectrum\textunderscore file}.

\subsubsection{Results}

\begin{figure*}
\begin{center}
\captionsetup{justification=centering}
  \begin{subfigure}[b]{0.45\textwidth}
    \includegraphics[trim={0 0 0 0}, clip,width=\textwidth]{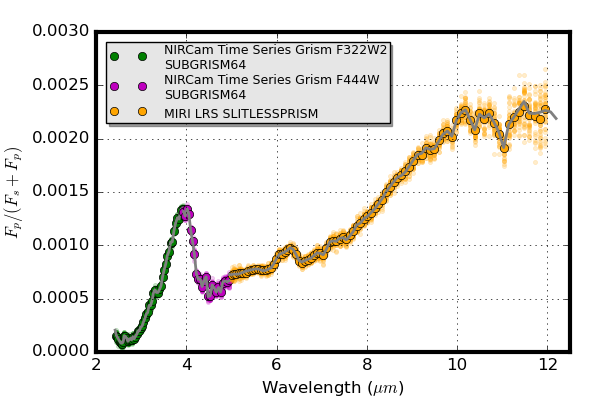}
    \caption{ }
    \label{fig:1}
  \end{subfigure}
  \begin{subfigure}[b]{0.45\textwidth}
    \includegraphics[trim={0 0 0 0}, clip,width=\textwidth]{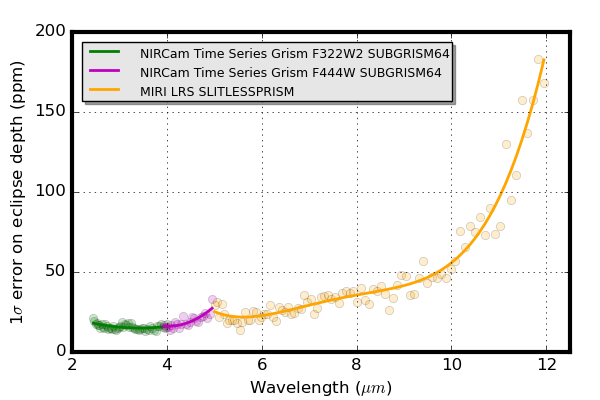}
    \caption{ }
    \label{fig:1}
  \end{subfigure}
  \begin{subfigure}[b]{0.80\textwidth}
    \includegraphics[trim={0 0 0 0cm}, clip,width=\textwidth]{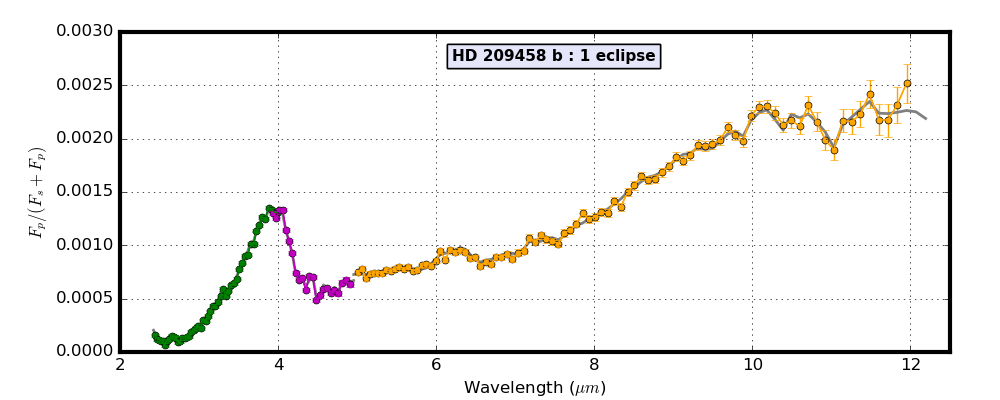}
    \caption{ }
    \label{fig:1}
  \end{subfigure}
\caption[]{Case 4: simulation of emission spectrum for HD 209458 b with NIRCam Time Series Grism F322W2 and F444W modes (both with SUBGRISM64) and MIRI LRS slitless mode, using the Monte Carlo method. a) Monte Carlo simulation recovered eclipse depths versus wavelength (large dots: average of 50 realisations, small dots: individual realisations, grey line:  input spectrum binned to $R$ = 100).
 b) 1$\sigma$ error on eclipse depth versus wavelength with 2nd and 4th order polynomial line fits for NIRCam and MIRI respectively. c) Synthetic post-processed dayside-emission spectrum for 1 eclipse (per instrument mode) (grey line: input spectrum binned to $R$ = 100). } 
\label{Fig:case4}
\end{center}
\end{figure*}

Figure \ref{Fig:case4} shows the results from this case study. We find that a very high precision result is obtained from just one eclipse per instrument mode given the brightness of the host star. Transit depths recovered from the Monte Carlo simulation are shown in Figure \ref{Fig:case4}a.  Figure \ref{Fig:case4}b shows the standard deviations for each bin and as previously polynomials are fitted to these (2nd order for NIRCam, 4th order for MIRI).

Using the polynomial fitted values, for NIRCam the eclipse depth precision ranges from 15 ppm (at 3.41 \textmu m) to 18 ppm (at 2.45 \textmu m) for the F332W2 mode, and from 16 ppm (at 4.06 \textmu m) to 27 ppm (at 4.95 \textmu m) for the 
F444W mode.  For MIRI LRS the precision ranges from a minimum of 22 ppm (at 5.64 \textmu m) to a maximum of 183 ppm (at 11.95 \textmu m). We can see that the randomised post-processed spectrum  has very small error bars and traces the input spectrum to high degree of precision and accuracy with just a single observation (per instrument mode). These precision results are comparable to those obtained in transmission for the same target.

\subsection{Case 5: Noise budget analysis of K2-18 b observation with MIRI using Allan deviation method}

\subsubsection{Overview}
A noise budget analysis breaks down the total noise into its constituent elements, helping us to understand which noise sources are most impactful and thus require the greatest attention for mitigation in observational and data reduction strategies.  Here we perform a noise budget for K2-18 b observed with MIRI LRS slitless mode. We use out-of-transit simulations with Allan deviation analysis, looping through different noise sources and recovering the fractional noise at $T_{14}$ versus wavelength for each source.   The noise sources are those listed in Table \ref{table: summary of noise table}. Each is simulated with the other noise sources switched off and a simulation is also produced with all noise sources on.
The instrumental effects listed in Table \ref{table: summary of noise table} (excluding the user-defined systematic) are applied in all cases with the exception of pointing jitter which is only applied in the pointing jitter noise case and in the all noise case.  Background sources listed in Table \ref{table: summary of noise table} are applied only if the corresponding noise is being simulated, e.g. the zodical light background is not applied unless the zodi noise is being simulated, so only in the zodi noise and all noise cases.  The results are processed through stage 1 and stage 2 (Allan deviation analysis) of the pipeline (Figure \ref{Fig:architecture}) with some variations.  The jitter decorrelation step is only applied in the pointing jitter and all noise cases.  Background subtraction is only applied if a diffuse background (zodi, optical surfaces emission or sunshade) is included.  Dark current subtraction is only applied for dark current noise and all noise cases. We obtain the average result of 10 realisations in each case.

\subsubsection{Simulation settings}
The simulation settings for the noise budget simulation are given in Table \ref{tab: simulation settings} and mostly follow those of case 1 with the following exceptions. The parameter \texttt{sim\textunderscore mode} is set to 3, to select the noise budget option.  When this is done \texttt{sim\textunderscore noise\textunderscore source} can be left blank as the code will loop through all noise sources. \texttt{obs\textunderscore inst\textunderscore config} is set for MIRI LRS SLITLESSPRISM. 
Also the auto-aperture sizing function is disabled (\texttt{pipeline\textunderscore auto\textunderscore ap} is set to 0) so that an extraction aperture size must be applied manually. This ensures that variation of the extraction aperture size is not a factor when comparing the different noise contributions. We set \texttt{pipeline\textunderscore ap\textunderscore factor} to 1 (based on the value obtained by the auto-aperture function for a simulation with all noise activated).  The width of the mask is
$2 \times \alpha F \lambda$, where $F$ is the F-number, $\lambda$ is the wavelength and $\alpha$ is the value of  \texttt{pipeline\textunderscore ap\textunderscore factor}. For a rectangular aperture mask (as selected here) $\lambda$ is the longest wavelength in the wavelength range. If a wavelength-dependent aperture mask is chosen, $\lambda$ is the wavelength of each pixel column.
 
\subsubsection{Results}

\begin{figure*}
\begin{center}
\captionsetup{justification=centering}
    \includegraphics[trim={1cm 0.5cm 0.5cm 1cm}, clip,width=1.0\textwidth]{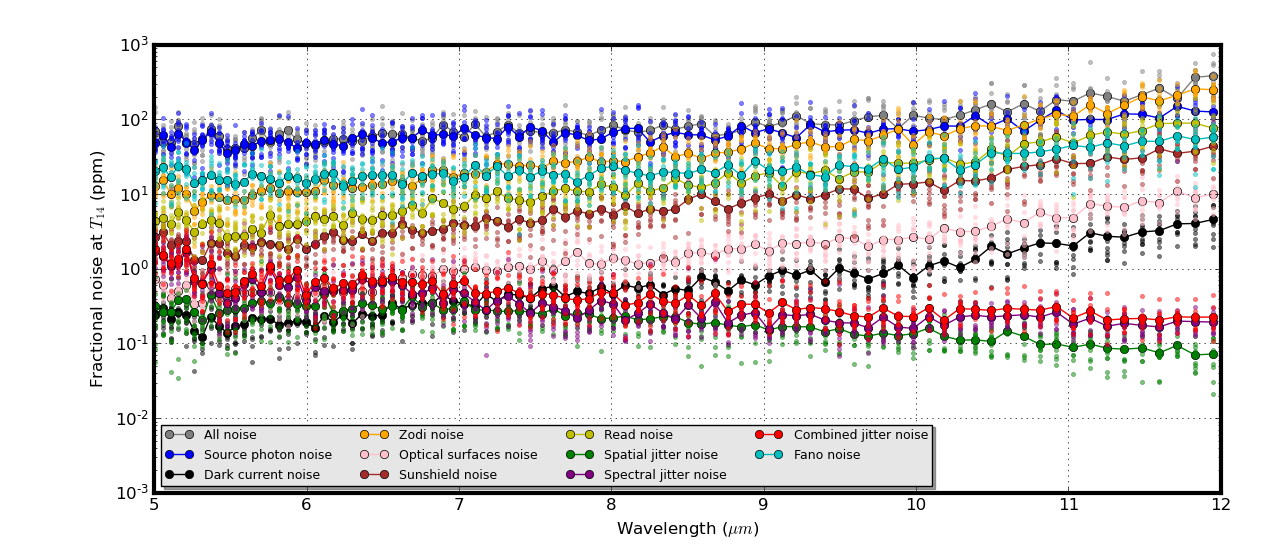}
\caption[]{Case 5: noise budget for MIRI LRS observing K2-18 b at $R$ =  100 (large dots: average of 10 realisations, small dots: individual realisations).}
\label{Fig:ex4}
\end{center}
\end{figure*}

Results from this example are shown in Figure \ref{Fig:ex4}. The smaller dots show results from individual realisations with the larger dots giving the average per spectral bin.  
We find that below about 10 \textmu m the spectra can be considered photon-noise-limited (the stellar `source photon noise' being the largest noise contributor), however at longer wavelengths zodiacal light noise becomes significant and exceeds source photon noise above $\sim$ 11 \textmu m.  Read noise is also significant at the long wavelength end but is mitigated by reading up-the-ramp and the large number of groups per integration (in this case $n=35$). Fano noise, which is dependent on the photons from both star and backgrounds, also becomes a significant noise source at the red end, bolstered by the high number of zodi photons, but does not exceed the source photon noise.  Noise from the sunshade emission increases with wavelength and is more significant than noise emission from optical surfaces, however it too does not exceed the source photon noise up to 12 \textmu m.
Other noise sources such as pointing jitter, optical surfaces emission noise, and dark current noise are well-controlled and orders of magnitude smaller than the source photon noise.  The contribution of some noise sources is dependent on the extraction aperture size. For example if we use a larger aperture size, the impact of zodi noise increases overtaking photon noise at shorter wavelengths. 
This kind of analysis can also be used to assess the impact of changing observational parameters such as saturation limit (and thus integration time and total number of integrations) on the overall noise balance.

\section{Conclusions}
JexoSim 2.0 is a time-domain simulator of JWST exoplanet spectroscopy designed for the planning and validation of science cases for JWST.  JexoSim generates synthetic spectra that capture the full impact of complex noise sources and systematic trends, allowing for assessment of both accuracy and precision in the final spectrum.
In this paper, we described the upgraded simulator, which has new noise sources and instrumental effects and  has been designed to improve operability, speed, efficiency and versatility.  JexoSim 2.0 is now freely available to the community.  We have described here how to implement the code, with specific case  studies  that can be replicated by a user.  

The case studies demonstrate a range of capabilities with simulated results from all four instruments and different subarray configurations.
We examined how JWST can be used to observe the hot Jupiter HD 209458 b around a Sun-like star, and the low-mass planet K2-18 b around an M-dwarf. Observing bright targets, such as  HD 209458 b, may presents challenges in finding subarray modes that will not saturate, with possible gaps in wavelength coverage when combining certain NIRSpec grating configurations.
K2-18 b  is of great current interest given its location in the habitable zone with similar insolation to the Earth and the discovery of water vapour and clouds in its atmospheric spectrum using the Hubble WFC3 IR instrument \citep{Benneke_2019, Tsiaris2019} using nine transits. However whereas the Hubble spectrum covered only 1.1 to 1.7 \textmu m, the JWST spectra simulated here range from $\sim$  0.9 to 12 \textmu m.  While we have not performed a spectral retrieval study in this paper, spectral features in the WFC3 range are clearly visible over noise with just one transit observation using NIRSpec or NIRISS. Similar levels of noise are obtained with each.  The 1.4 \textmu m water feature (observed with the Hubble WFC3) has an amplitude of $\sim$  115 ppm in the spectrum from our radiative transfer model.  JexoSim predicts a transit depth noise of 33 ppm with NIRSpec (G140M) and 31 ppm with NIRISS. If we divide the amplitude by the transit depth noise we obtain a signal-to-noise of about  3.5-3.7 for one transit, which compares well with the multiple transits needed to achieve a similar precision with Hubble. 
When observing K2-18 b, NIRCam achieves similar noise  performance to NIRSpec and NIRISS, though its coverage does not reach down to the WFC3 wavelength range.  
 Whilst NIRSpec offers the widest total wavelength coverage of the three NIR instruments, when using the gratings, this must be achieved through applying multiple subarray modes, each requiring its own observation. The PRISM mode can cover the complete range in one pass, however this will work only with dim targets that will not saturate the detector.
For K2-18 b, the noise at $R$=100 with MIRI LRS ranges from 70 ppm to 448 ppm. The noise increases with wavelength, so that  particularly at the long wavelength end of MIRI, these results may not be useful.  However, binning to lower $R$ will improve precision.
Future studies using spectral retrievals can incorporate JexoSim's ability to add confounding factors such as spot effects or additional systematics to see if spectral features including simulated biomarkers can be successfully recovered from low-mass planets around M-dwarf stars, targeted for observation by JWST.

Although there may be systematics not currently known or included in JexoSim 2.0, the code can be easily upgraded or modified in future to accommodate these.  JexoSim 2.0 can be used to cross-validate initial results from other simulation tools in cases where time-domain effects are critical to consider, as well as compare performance across the suite of JWST instruments and modes as performed here.  JexoSim 2.0 provides a framework for investigating in detail the feasibility of specific and novel scientific cases for JWST where detectability may be vulnerable to astrophysical or instrumental confounding factors.  It is also well-suited to investigating the impact of star spots and faculae as well as stellar pulsation and granulation noise.  With simple modifications it can be used to investigate detection of time-dependent scientific signals such as exomoon transits or phase-related modulations. JexoSim 2.0 can also be customised for other astrophysical applications beyond exoplanet spectroscopy, such as brown dwarf spectroscopy.

With JWST, exoplanet science will be transformed.  JexoSim 2.0 will assist in maximising and optimising the scientific possibilities of this ground-breaking observatory.

\section*{Acknowledgements}
Thanks to Sam Rowe (Cardiff University) for alpha testing and help with implementing Numba. Thanks to Andreas Papageorgiou (Cardiff University) for alpha testing and to Tyler Gordon (University of Washington) for trialling the code and providing valuable feedback. Thanks to Jayesh Goyal (Cornell University) for help with using the ATMO scaleable grid. We use a modified version of his publicly-available code snippet for rescaling the spectra from this grid for use in JexoSim. Our thanks to the referee for their helpful comments and suggestions for improvement.
S.S. was supported by UKSA grant ST/S002456/1. 

\section*{Data Availability Statement}
The data underlying this article will be shared on reasonable request to the corresponding author.



\bibliographystyle{mnras}
\bibliography{example.bib} 


\bsp	
\label{lastpage}
\end{document}